\documentclass{aa}  

\usepackage{graphicx}
\usepackage{txfonts}
\usepackage[usenames, dvipsnames]{color,colortbl}

\usepackage[colorlinks=true, allcolors=blue, draft]{hyperref}

\usepackage{amsmath}
\usepackage{amssymb}
\usepackage[colorinlistoftodos]{todonotes}

\newcommand{\Eq}[1]{Eq.~\ref{#1}}

\newcommand{\Fig}[1]{Fig.~\ref{#1}}

\newcommand{\Ntotal}{494}
\newcommand{\Nsamp}{182}
\renewcommand{\vec}[1]{\mathbf{#1}}
\newcommand{\Ntng}{158}
\newcommand{\Nmock}{790}
\newcommand{\SNRmin}{3}

\def\editorial{true}

\ifdefined\editorial
    \newcommand{\nb}[1]{{\color{red}TODO:#1}}
\else
    \newcommand{\nb}[1]{{}}
\fi

\newcommand{\sggg}[1]{\textcolor{green}{[]}}

\def\rt{R_{\rm t}}
\def\Re{R_{\rm e}}
\def\Rhalf{\Re} 
\def\js{\tilde j_{\star}}
\newcommand{\FITgamma}{0.65^{+0.06}_{-0.08}}
\newcommand{\FITzalpha}{-0.27^{+0.42}_{-0.56}}
\newcommand{\FITalpha}{0.03^{+0.01}_{-0.01}}
\newcommand{\FITbeta}{0.37^{+0.09}_{-0.05}}
\newcommand{\FITZP}{2.79^{+0.05}_{-0.03}}

\newcommand{\HI}{\hbox{{\rm H}{\sc \,i}}}

\def\Ha{{\rm H\alpha}}

\def\AREPO{{\small AREPO}}

\def\galpak{{\textsc GalPaK$^{\rm 3D}$}}
\def\sersic{{S{\'e}rsic}}

\newcommand{\OII}{\hbox{[{\rm O}{\sc \,ii}]}}
\def\log{{\rm\thinspace log}}

\newcommand{\Mvir}{M_{\rm vir}}
\newcommand{\Mh}{M_{\rm h}}
\newcommand{\jh}{j_{\rm h}}

\def\dex{{\rm\thinspace dex}}
\def\cm{{\rm\thinspace cm}}

\def\pc{{\rm\thinspace pc}}
\def\kpc{{\rm\thinspace kpc}}

\def\Mpc{{\rm\thinspace Mpc}}

\def\kms{{\rm\thinspace km\thinspace s}^{-1}}

\def\Msun{\hbox{$\rm\thinspace M_{\odot}$}}

\def\yr{{\rm\thinspace yr}}

\def\Gyr{{\rm\thinspace Gyr}}

\def\Msunpc2{{\Msun\pc}^{-2}}
\def\Msunyrkpc2{{\Msun\yr^{-1}\kpc}^{-2}}
\def\arcsec{{\rm\thinspace arcsec}}
\def\magarcsec2{{\rm\thinspace mag\thinspace arcsec}^{-2}}

\begin{document}

   \title{The MUSE Hubble Ultra Deep Field Survey XVI. The angular momentum of low-mass star-forming galaxies. A cautionary tale and insights from TNG50\thanks{This study is based on observations collected at the European Southern
Observatory under ESO programmes 094.A-0289, 095.A-0010 and 096.A-0045. }}


   \author{Nicolas  F. Bouch\'e 
          \inst{1} 
              \and
          Shy Genel\inst{2,3}
          \and
          Alisson Pellissier\inst{4,5}
          \and
          Cédric Dubois\inst{4}
          \and
          Thierry Contini\inst{4}
          \and
          Beno\^it Epinat\inst{6}
          \and
          Annalisa Pillepich\inst{7}
          \and
          Davor Krajnovi\'c\inst{8}
          \and
          Dylan Nelson\inst{9}
          \and
          Valentina Abril-Melgarejo\inst{6}
          \and
           Johan Richard\inst{1}
          \and
         Leindert Boogaard\inst{7,10}
          \and
          Michael Maseda\inst{10,11}	
          \and
          Wilfried Mercier\inst{4}
          \and
          Roland Bacon\inst{1}
          \and
          Matthias Steinmetz\inst{8}
          \and
          Mark Vogelsberger\inst{12}
          }

   \institute{Univ Lyon, Univ Lyon1, Ens de Lyon, CNRS, Centre de Recherche Astrophysique de Lyon (CRAL) UMR5574, F-69230 Saint-Genis-Laval, France\\
              \email{nicolas.bouche@univ-lyon1.fr}
         \and
         Center for Computational Astrophysics, Flatiron Institute, 162 Fifth Avenue, New York, NY 10010, USA
         \and
         Columbia Astrophysics Laboratory, Columbia University, 550 West 120th Street, New York, NY 10027, USA
     \and
     Institut de Recherche en Astrophysique et Plan\'etologie (IRAP), Universit\'e de Toulouse, CNRS, UPS, F-31400 Toulouse, France
     \and 
      Laboratoire Lagrange, Université Côte d’Azur, Observatoire de la Côte d’Azur, CNRS, Blvd de l’Observatoire, F-06304 Nice cedex 4, France
      \and
    Aix Marseille Univ, CNRS, CNES, LAM, Marseille, France
    \and
    Max-Planck-Institut f\"ur Astronomie, K\"onigstuhl 17, 69117 Heidelberg, Germany 
    \and
     Leibniz-Institut für Astrophysik Potsdam (AIP), An der Sternwarte 16, D-14482 Postdam, Germany
     \and
    Universit\"{a}t Heidelberg, Zentrum f\"{u}r Astronomie, Institut f\"{u}r theoretische Astrophysik, Albert-Ueberle-Str. 2, 69120 Heidelberg, Germany
     \and
     Leiden Observatory, Leiden University, PO Box 9513, NL-2300 RA Leiden, The Netherlands
      \and
     Department of Astronomy, University of Wisconsin-Madison, 475 N. Charter Street, Madison, WI 53706, USA
       \and
     Department of Physics, Massachusetts Institute of Technology, 77 Massachusetts Avenue, Cambridge, MA 02139, USA
          }

   \date{Received---; accepted ---}

    \titlerunning{The angular-momentum of low-mass galaxies}
     \authorrunning{Bouché et al.}
 
\abstract{
We investigate the specific angular momentum (sAM) $ j(<r)$ profiles of intermediate redshift ($0.4<z<1.4$) star-forming galaxies (SFGs)
in the relatively unexplored regime of low masses (down to $M_\star\sim10^8\Msun$) and small sizes (down to $\Re\sim 1.5$ kpc), and we characterize the sAM scaling relation (i.e., Fall relation) and its redshift evolution.
We have developed a 3D methodology to constrain sAM profiles of the star-forming gas using a forward modeling approach with \galpak{} that incorporates the effects of beam smearing, yielding the intrinsic morpho-kinematic properties even with limited spatial resolution data.
Using mock observations from the TNG50 simulation,
we find that our 3D methodology robustly recovers the star formation rate (SFR)-weighted $\js(<r)$ profiles down to a low effective signal-to-noise ratio (S/N) of $\gtrapprox3$.
We  applied our methodology blindly to a sample of \Ntotal\ \OII{}-selected SFGs in the MUSE Ultra Deep Field (UDF) 9~arcmin$^2$ mosaic data, covering the unexplored $8<\log M_\star/\Msun<9$ mass range. We find that the (SFR-weighted) sAM relation follows $\js\propto M_\star^{\alpha}$  with an index $\alpha$ varying from $\alpha=0.3$ to $\alpha=0.5$, from $\log M_\star/\Msun=8$ to $\log M_\star/\Msun=10.5$. 
The UDF sample supports a redshift evolution $\js\propto (1+z)^a$, with $a={\FITzalpha}$ which is consistent with the  $(1+z)^{-0.5}$ expectation from a universe in expansion. The scatter of the sAM sequence is a strong function of the dynamical state  with $\log j|_{M_\star}\propto \FITgamma \times \log(V_{\rm max}/\sigma)$, where $\sigma$ is the velocity dispersion at $2\Re$.
In TNG50,  SFGs also form a $\js-M_{\star}-(V/\sigma)$ plane, but it correlates more with galaxy size than with morphological parameters. Our results suggest that SFGs might experience a dynamical transformation, and lose their sAM, before their morphological transformation to becoming passive via either merging or  secular evolution.
}

\keywords{
galaxies: high-redshift
galaxies: evolution 
galaxies: kinematics and dynamics
galaxies: structure
}

\maketitle
 

\section{Introduction}

In a $\Lambda$ cold dark matter ($\Lambda$CDM) universe, baryons cool, fall inwards, and form centrifugally supported disks in the centers of halos.
The specific angular momentum (sAM) 
sets the pressure, instabilities, gas fractions \citep{ObreschkowD_16a,RomeoA_18a,RomeoA_20b,LiJ_20a}, and most importantly determines the disk properties, such as size  \citep[e.g.,][]{WhiteS_78a,FallM_80a,FallM_83a,MoH_98a,DalcantonJ_97a,vandenBoschF_03b,DuttonA_09a,SomervilleR_17a}.
As disks evolve from $z=2$ to the present, they must grow from
a vast reservoir of corotating cold accreting material
in the circum-galactic medium (CGM) as argued in \citet{RenziniA_20a}.
This is also strongly supported by
hydro-dynamical simulations 
which predict roughly coplanar gaseous structures \citep{StewartK_13a,StewartK_17a,DanovichM_15a,HoS_19a,KretschmerM_20a,DeFelippisD_21a} embedded in a rotating CGM
\citep{DeFelippisD_20a}. {
These coplanar structures were initially found by  \citet{BarconsX_95a} and \citet{SteidelC_02a} in a handful of  background quasar sight-lines,
but they are now routinely observed  on scales of 20-80 kpc
\citep[as in][]{BoucheN_13a,BoucheN_16a,HoS_17a,LopezS_18a,MartinC_19a,ZablJ_19a}. These structures are important as they not only
bring fuel for star-formation but also  angular momentum \citep{RenziniA_20a,DeFelippisD_21a}.
}

In this context, one of the most fundamental properties of disk galaxies is their sAM \citep[as argued in][]{FallM_80a}.
The sAM of halos ($\jh=J_{\rm h}/\Mh$) is tightly correlated to the halo mass $\Mh$ ($\jh\propto \Mh^{2/3}$)
as a result of tidal torques from the large-scale structure  \citep[e.g.,][]{PeeblesP_69a,EfstathiouG_79a}.
It is often assumed that the sAM of baryons and of dark matter are equal at accretion, and that as disks form in the centers of halos the baryonic sAM is conserved \citep[e.g.,][]{FallM_80a,MoH_98a,BurkertA_10a}.
While this assumption of $j$ conservation during collapse has been appealing since the late 1970s \citep[e.g.,][]{WhiteS_78a,FallM_80a}, it is not trivial given the infalling baryons can both lose and gain angular momentum from the virial radius to the inner disk \citep[e.g.,][]{NavarroJ_00a,SharmaS_12a,DanovichM_15a,GenelS_15a,DeFelippisD_17a,JiangF_19a}.
In particular, \citet{GenelS_15a} and \citet{DeFelippisD_17a} found that galactic winds are essential for producing late-type galaxies with sufficient angular momentum.
%


Observationally, the stellar sAM of massive disks (with $\log M_\star/$M$_\odot>9.5$) follows a
 similar scaling relation as the dark matter $\jh-\Mh$ scaling relation  with $j_\star\propto M_\star^{\alpha}$ with $\alpha\simeq0.6$,
 as  reported by M. Fall and collaborators 
\citep[][hereafter FR13]{FallM_83a,RomanowskyA_12a,FallM_13a} and recently revisited by  \citet[][hereafter FR18]{FallM_18a}.
This  similarity between the halo and baryonic sAM scaling relations, both in  normalization (for galaxy disks) and slope, highlights a potential intimate link between galaxies and their host  DM  haloes, as discussed extensively in \citet{PostiL_18a}. However, hydrodynamical simulations tend to show a more complex relation between the DM and baryonic sAM \citep[e.g.,][]{GenelS_15a,JiangF_19a}.

Since the pioneering study of \citet{FallM_83a},  the $j_\star-M_\star$ relation has become a standard tool in galaxy evolution research
thanks to recent  Integral Field Spectrograph (IFS)  surveys
 and to more accurate  simulations  of disk formation  \citep[e.g.,][]{VogelsbergerM_12a,VogelsbergerM_13a,SchayeJ_15a,CeverinoD_17a,GrandR_19a,NelsonD_19a,PillepichA_19a}, see review in \citet{VogelsbergerM_20a}.
 Hence, understanding the slope and normalization of 
this important scaling relation is of paramount importance as it is intimately related to galaxy morphological diversity and holds clues to accretion of baryons onto galaxies \citep{GenelS_15a,TekluA_15a,ZavalaJ_16a,LagosC_17a,ElBadryK_18a}.

In the local universe, several groups have expanded on the work of Fall and collaborators
to confirm  the $j_\star-M_\star$ relation with increasingly large samples of galaxies 
as in \citet{CorteseL_16a} and \citet{LapiA_18a} which each had $\sim500$ galaxies, 
albeit mostly with   $M_\star>10^{9.5} \Msun$.
Some of these local surveys have shown that the scatter of the $j_\star-M_\star$ relation 
 strongly correlates  with morphology 
\citep{ObreschkowD_14a,CorteseL_16a,RizzoF_18a,SweetS_18a}.
At lower masses $M_\star<10^9 \Msun$, the $j_\star-M_\star$ relation is not understood  as well either theoretically \citep[e.g.,][]{StevensA_16a,MitchelP_18a,ElBadryK_18a}
or observationally \citep[but see][]{ButlerK_17a,PostiL_18b,ManceraPinaP_21a}. 
 Both the sample of 14 dIrr galaxies in the $10^6$--$10^9~\Msun$ mass range from \citet{ButlerK_17a}
 and the  local disks with $M_\star>10^7\Msun$  from \citet{PostiL_18b} and \citet{ManceraPinaP_21a}
 show that the $j_\star-M_\star$ relation is well described by a single power-law at all scales {with a slope $\alpha=0.6\pm0.1$ \cite[see][]{FallM_18a}.}
 
At high redshifts, studying the $j-M$ relation has  become particularly important in order to constrain the redshift evolution of this fundamental scaling relation. Thanks to recent advancements with second generation IFS instruments, such
as the Multi-Unit Spectroscopic Explorer (MUSE; \citealp{BaconR_10a}) and the
K-band Multi-Object Spectrograph (KMOS; \citealp{SharplesR_13a}),
several groups have constrained the $j-M$ relation using increasingly larger
samples of $z>1$ star-forming galaxies (SFGs). Among these, there is the sample of 360 SFGs from the SINS/KMOS3D survey  \citep{BurkertA_10a,BurkertA_16a,WisnioskiE_19a} at redshifts $0.8<z<2.3$, the sample of $\approx400$ SFGs at $z\sim0.8$ of
\citet{SwinbankM_17a} collected in MUSE and KMOS surveys, and the sample of 486 $z=0.6-1.0$ SFGs \citep{HarrisonC_17a} from the KROSS  $\Ha$ survey \citep{StottJ_16a}.
These surveys \citep[and others, e.g.~][]{AlcornL_18a,GillmanS_20a}
are often limited to relatively massive SFGs  with a typical limit of $M_\star>10^{9.5}\Msun$.
 
The $j-M$ relation has also not been explored in the low mass regime at $M_\star$ of $10^8\Msun$ to $10^{9.5}\Msun$ beyond the local universe \citep{PostiL_18b}.
Thanks to the MUSE IFS \citep{BaconR_10a} and its exquisite sensitivity, it is now possible to study the kinematics of SFGs at low masses down to $M_\star=10^8\Msun$, as first demonstrated by \citet{ContiniT_16a}.
The low-mass  regime ($M_\star=10^{8-9}\Msun$) is challenging as galaxies become marginally resolved (when smaller than 3~kpc), and it becomes difficult to recover reliable measurements of both sizes and rotation velocity.

This paper aims to show that the $j-M$ relation can be characterized in small and low-mass galaxies at redshifts from $z=0.4$--1.4, opening up new parameter space thanks to the MUSE instrument and to our  3D modeling approach. {Contrary to galaxies in the local universe where the $j-M$ relation can be studied independently for the stellar $j_\star$ or gas  components $j_{\rm gas}$  \citep[e.g.,][]{ObreschkowD_14a,ObreschkowD_16a,LiJ_20a,ManceraPinaP_21a}, this is currently extremely challenging in distant galaxies at $z\simeq1.0$ or beyond.
One should note that the gas sAM $j_{\rm gas}$ in this context is derived solely from $\HI$ data at $z=0$, and it is often assumed that the sAM for the star-forming gas (molecular or ionized) is similar to $j_\star$ because of the similar extent \citep{LeroyA_08a,NelsonE_16a,WilmanD_20a} and kinematics \citep{MartinssonT_13b,LangP_20a}~\footnote{See \citet{MarascoA_19a} for details on the conversion between $j_{\rm gas}$ and $j_\star$.}.
As discussed in the next section, we focus our analysis on the sAM $j$ derived from the star-forming gas, denoted $\js$, both for a large sample of \Ntotal\ SFGs  from the MUSE {\it Hubble} Ultra Deep Field (UDF) observations presented in \citet{BaconR_17a} and 
for SFGs   from the TNG50 simulations \citep{NelsonD_19a,PillepichA_19a}.
}

This paper is organized as follows.
In section~\ref{section:methodo}, we present our methodology that is designed to be more accurate in the low-mass regime.
In section~\ref{section:Illustris}, we validate our methodology on 158 simulated galaxies taken from TNG50.
In section~\ref{section:UDF}, we apply our methodology on \Ntotal\ galaxies in the MUSE UDF. 
In section~\ref{section:tng50:comparison}, we compare our results to $\sim5000$ SFGs in TNG50.
 We discuss the implication of our results in section~\ref{section:implications}.
Finally, we present our conclusions in section~\ref{section:conclusions}.
 Throughout this paper, we use a ``Planck 2015''  cosmology \citep{Planck2015} with $\Omega_M=0.307$, $\Lambda=0.693$, $H_0=67$ km/s/Mpc, yielding 8.23 kpc/arcsec, and we
consistently use ``log'' for the base-10 logarithm.

\section{Measuring angular momentum}
\label{section:methodo}

\subsection{Definitions}

The sAM $j\equiv J/M$ is, in general, given by
\begin{eqnarray}
j_{\rm tot}&\equiv& \frac{J}{M}=\frac{\int \rho(\vec{x}) (\vec{v}\times \vec{x}) {\rm d}^3\vec{x}}{\int \rho(\vec{x}) {\rm d}^3\vec{x}} \label{eq:AM:3d}.
\end{eqnarray}
In practice, the two-dimensional version of the specific angular momentum, appropriate for an axi-symmetric disk,
is used
\begin{eqnarray}
j(<R)=j_{\rm 2d}(<R)&=&\frac{\int_0^{R} \Sigma({r}) {v}(r){ r^2}\, {\rm d}{r}}{\int_0^{R}\Sigma({r}) r\,{\rm d}{r}}\label{eq:AM:2d},
\end{eqnarray}
where $R$ is the enclosed radius, $\Sigma(r)$ the surface brightness, and $v(r)$ the rotation velocity profile.

Regarding Eq.~\ref{eq:AM:2d}, 
the total angular momentum $j_{\rm tot}\equiv j(<\infty)$  is not observable directly as inherent surface brightness limitations preclude measurements beyond $R_e$ or at best 2$R_e$ \citep[e.g.,][]{MarascoA_19a}. For massive galaxies, probing $j$ up to $R_e$ or 2$R_e$, where the $j$ profile reaches its plateau, might be sufficient, but in low mass galaxies, that regime might be difficult to reach even in the local universe, as discussed in \citet{PostiL_18b,MarascoA_19a}. Thus, previous work often resorted to estimators for $j_{\rm tot}$ such as \Eq{eq:RF12}, which is discussed in \S~\ref{section:methodo:RF12}.
Here, we  use a 3D methodology (\S~\ref{section:methodo:3D})  that bypasses common approximations { and similar in essence to the analysis of \citet{ManceraPinaP_21a} who used $^{\rm 3D}$Barolo \citep{DiTeodoroE_15a}.}
In \S~\ref{section:Illustris}, we show with mock cubes from TNG50 that our 3D method combined with Eq.~\ref{eq:AM:2d} does reproduce the true $j$ values, computed directly from Eq.~\ref{eq:AM:3d}.

It is worth noting that in order to determine $j_\star(r)$ from Eq.~\ref{eq:AM:2d}, one needs the stellar mass profile $\Sigma_\star(r)$ and the stellar rotation curve $v_\star(r)$.
However, determining stellar kinematics beyond $z=0$ is challenging in SFGs \citep[but see][for a first attempt]{GuerouA_17a} or even in passive galaxies with strong continuum \citep[e.g.,][]{vandeSandeJ_13a,BezansonR_18a,ColeJ_20a}.
As a result, most studies  of the $j_\star-M_\star$ relation use the star-forming gas $v_{\Ha}$ kinematics as a proxy for $v_\star$ \citep[e.g.,][]{BurkertA_16a,HarrisonC_17a,SwinbankM_17a}.
Two studies attempted to convert 
$v_{\Ha}$ to $v_\star$, \citet{MarascoA_19a} at $z=1$ and \citet{PostiL_18b} at $z=0$ using  the asymmetric drift correction \citep{MeurerG_96a,DalcantonJ_10a,BurkertA_10a}.

Regarding the stellar mass profile $\Sigma_\star(r)$, it acts as the normalization factor in Eq.~\ref{eq:AM:2d}, and thus one only need to know the profile shape, namely the Sersic index and the half-light radius $R_e$.
For SFGs, one can use the star-forming gas profile  $\Sigma_{\Ha}$ (or $\Sigma_{\OII}$) as a proxy for  $\Sigma_\star$ provided that the SF gas and stars have similar scale length~\footnote{{This implicitly assumes a small or negligible bulge component.}}, which is expected from the almost linear Kennicut-Schmidt relation \citep{KennicuttR_98a,GenzelR_10a,DaddiE_10b,BacchiniC_19a}.
In fact, the $\Ha$ size $R_{\rm e, \Ha}$ to continuum size ratio $R_{\rm e,\star}$ has been found to be $\approx1$
\footnote{This ratio is slightly mass-dependent beyond $10^{10}\Msun$ reaching 1.2-1.3 at $10^{11}\Msun$ in the KMOS3D survey \citep{WilmanD_20a}. } from a very large sample of 3200 galaxies at $0.7<z<1.5$ with WFC3 imaging in \citet{NelsonE_16a}, confirmed in our data as discussed in \S~\ref{section:udf:sizes}. 

In this paper,
we measure the  star formation rate (SFR)-weighted sAM $j_{\rm sfr}$ determined from the star-forming gas properties, namely $v_{\OII}(r)$ and $\Sigma_{\OII}(r)$,
as a proxy for $j_{\star}$ following other studies \citep[e.g.,][]{BurkertA_16a,ContiniT_16a,SwinbankM_17a}. 
This has the advantage that the measurement is self-contained (from $\Ha$ or $\OII$) and that a comparison to similar quantities from simulations  can be performed. 
We  use the notation $\js$ to designate $j_{\rm sfr}$ and
in Section~\ref{section:tng50:results}, we  quantify the potential bias between $\js$ and $j_\star$ in the TNG50 galaxies. 

\subsection{2D Methodology: An Approximate sAM}
\label{section:methodo:RF12}

Given the difficulties in measuring $j(<r)$ in the outskirts of galaxies,
a method that is commonly used in the literature \citep[e.g.,][]{BurkertA_16a,SwinbankM_17a,HarrisonC_17a,GillmanS_20a} 
for estimating $j_{\rm tot}$ is the  following formulae \citep[][hereafter RF12]{RomanowskyA_12a}:
\begin{equation}
j_{\rm tot, RF} = k_n \,v_s\, \Rhalf ,\label{eq:RF12}
\end{equation}
where $\Rhalf$ is the half-light radius, $v_s$ the rotation velocity at $2R_e$, $v(2\,R_e)$, and  $k_{n}=1.15 + 0.029\cdot n + 0.062\cdot n^2$ is a constant that depends on the \citet{SersicJ_63a} index $n$. 
 \Eq{eq:RF12} reduces to the well-known $j_{\rm tot, RF} =2\,v_s\, R_d$ for disks with $R_d$ being the exponential scale length.

However, we caution the reader against using Eq.~\ref{eq:RF12}  for low mass galaxies with $\log (M_\star/\Msun)<9.5$--10, for the following  reasons.
  \Eq{eq:RF12}  assumes that the velocity profile $v(r)$ is a  constant  from $r=0$, which might be
 appropriate for galaxies with $\log(M/M_{\star})>10$,  but is less suitable for intermediate and low mass galaxies,
 which are known to have slowly rising rotation curves \citep[e.g.,][]{PersicM_96a,CatinellaB_06a}.
Consequently, Eq.~\ref{eq:RF12} can lead to an over-estimation
  of the total angular momentum as discussed   in \citet{ObreschkowD_14a}.

\begin{figure*}
\centering
\includegraphics[width=8.5cm, height=7.cm]{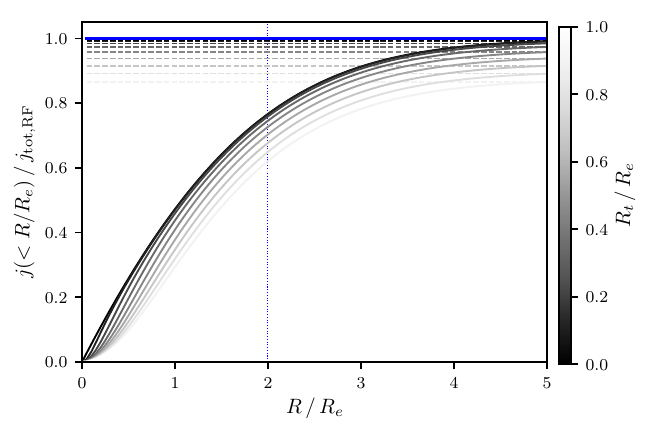}
\includegraphics[width=8cm]{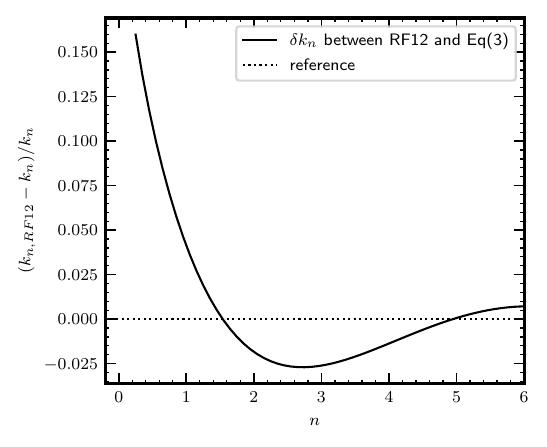}
\caption{Systematics at play in measuring $j(<r)$. {\bf Left:} The sAM profile $j(<r)$ normalized by the RF12 approximation (\Eq{eq:RF12}) as a function of radius  for disks with \sersic\ $n=1$ and a $\tanh$ rotation curve. The solid blue horizontal line represents the \Eq{eq:RF12} approximation. The solid curves represent the sAM for various rotation curves with various turn-over radii $\rt$ where the opacity of the line is inversely proportional to $\rt/\Re$. The vertical dotted line represents $2\Re$.
For galaxies with slowly rising rotation curves, with larger $\rt$, the RF12 approximation leads to a large overestimation of the total angular momentum (represented by the horizontal dashed lines), by 10 to 20\%. 
{\bf Right:} The relative difference for the $k_n$ coefficient between the $k_n$ formula from RF12 and \Eq{eq:kn:main} as a function of \sersic\ index $n$.
One sees that \Eq{eq:RF12} under- or over-estimate $k_n$ by up to $\sim10$\% for $0.5<n<4$. 
}
\label{fig:RF12pb}
\end{figure*}

  To quantify the overestimation in the low-mass regime, we show in
 Fig.~\ref{fig:RF12pb}(a)  several sAM profiles (solid lines), corresponding to ``tanh'' rotation curves, $v(r)=V_{\rm max}\tanh(r/\rt)$ with varying inner slopes (parameterized by the turn-over radius $\rt$), normalized to the RF12 approximation (represented by the blue horizontal line).
This figure shows that, for galaxies with slowly rising rotation curves, that is with $\rt/R_e$ of order unity,   Eq.~\ref{eq:RF12} results in an overestimation of the angular momentum by up to 15\%. \citet{ObreschkowD_14a} made a similar point using rotation curves $v(r)\propto(1-\exp(-r/\rt))$.

Moreover,  there can be significant differences between the often used $k_n$ approximation, namely $k_{n}=1.15 + 0.029\cdot n + 0.062\cdot n^2$  and the exact expression (derived in Appendix~\ref{sec:appendix:analytic}) given by:
\begin{equation}
\label{eq:kn:main}
k_n = \left(\frac{1}{b_n}\right)^{1/n}\frac{\Gamma(3n)}{\Gamma(2n)}
\end{equation}
that can be as large as 15\%\ for \sersic\ indices  $n\lesssim1$ as indicated in \Fig{fig:RF12pb}(b).

 In conclusion, combining these two effects, Eqs.~\ref{eq:RF12} and ~\ref{eq:kn:main},   can lead to an overestimation of the sAM (up to 20--25\%) for galaxies with $n\lesssim1$ and/or when the rotation curve is slowly rising, that is in the low-mass regime at $M_\star<10^{9.5}\Msun$.
In addition, such galaxies with a large turn-over radius $\rt$ tend to have large sizes and hence a low surface brightness, and thus their true $V_{\rm max}$ is more difficult to measure beyond $\rt$.


\subsection{3D Methodology}
\label{section:methodo:3D}

In order to estimate $j$ directly from Eq.~\ref{eq:AM:2d}, one needs the {intrinsic} (deprojected and free from instrumental and resolution effects) velocity and surface brightness flux profile and these can be measured directly from 3D (position-position-velocity) data
 using the \galpak\ algorithm\footnote{Available at \url{http://galpak3d.univ-lyon1.fr}.} described in \citet{BoucheN_15a}.

Briefly, \galpak\  performs a parametric fit on emission line data for the morphology and kinematics simultaneously directly onto the 3D data using a 3D ($x,y,\lambda$) disk model which specifies
the morphology and kinematic profiles.
For the morphology, the model assumes a \citet{SersicJ_63a} surface brightness profile $\Sigma(r)$, with \sersic\ index $n$.
For the kinematics, the model assumes a parametric form for the rotation curve $v(r)$  and  for the dispersion $\sigma(r)$ profile. 
  The model parameters are the centroid position $x$, $y$, $z$, the total line flux, the inclination $i$, the major-axis position angle P.A, the half-light radius ($\Rhalf$), the turnover radius $\rt$  and maximum velocity $V_{\rm max}$ for the rotation curve, and the velocity dispersion $\sigma_{\rm 0}$, described below.

The rotation curve $v(r)$ can be   ``arctan'', `tanh' or other analytical forms as described in the documentation~\footnote{See \href{http://galpak3d.univ-lyon1.fr/doc/}{http://galpak3d.univ-lyon1.fr/doc/}.}.
  As described in \citet{BoucheN_15a}, the  velocity dispersion profile $\sigma(r)$ is made of three components 
  added in quadrature. The first component $\sigma_1$ is that of a thick disk \citep{GenzelR_08a}, namely $\sigma_1(r)/v(r) =h_z/r$ where $h_z$ is the disk thickness, taken to be $0.2\times\Re$. The second component $\sigma_2$ is comes from the natural broadning of the 3D disk model of finite thickness.
  The third component  $\sigma_{3}$ is  an adjustable constant term $\sigma_0$, as in
   \citet{GenzelR_08a, ForsterSchreiberN_06a,CresciG_09a,WisnioskiE_15a,UblerH_19a} and describes any additional turbulent component to the kinematic dispersions. 

\galpak\ convolves the model with the Point Spread Function (PSF) and the instrumental Line Spread Function (LSF), which implies that the parameters are  ``intrinsic'', that is corrected for beam smearing and instrumental effects.  Here, we use a Moffat PSF as  characterized in \citet{BaconR_17a}, and 
 we use Eq.~(7) of  \citet{BaconR_17a} for the LSF.
 
\galpak\ uses a simple Metropolis-Hasting (MH) algorithm to optimize the parameters whose posterior distributions are given by the chain posteriors. 
As described in \citet{BoucheN_15a}, the \galpak\ implementation of the MH uses a nonstandard Cauchy proposal distribution which shortens the burn-in phase considerably, but suffers, like  most MH algorithms, from the need to manually tune the width of the proposal distribution.

Since its initial release, the following improvements and new features have been implemented in \galpak. First, in order to quantify the chain convergence for each parameter, \galpak\ performs the \citet{GewekeJ_92a}'s convergence diagnostic test for each of the parameters, which is  a $Z$-test of equality of means (of two parts of the chain) where auto-correlation in the samples is taken into account. Second, \galpak\ can now adapt the widths of the Cauchy proposal distribution automatically.  
Third, and most importantly, \galpak\  is  now (since v1.20) compatible with the \textsc{emcee} code \citep{emcee}. \textsc{emcee} has the advantage that its multi-walkers remove the need to tune the proposal distribution, but it has the disadvantage to retain the Normal proposal distribution which is very slow and inefficient in reaching convergence because the Gaussian function does not have broad wings.
For our purposes, we combine the self-tuning of our fast MH algorithm implemented in \galpak\ with the \textsc{emcee} sampler once convergence is reached as follows: we first run the self-tuning of our fast MH algorithm with Cauchy sampling, and then use  the \textsc{emcee} sampler once convergence is reached.

\begin{figure}
 \centering
\includegraphics[width=0.8\linewidth]{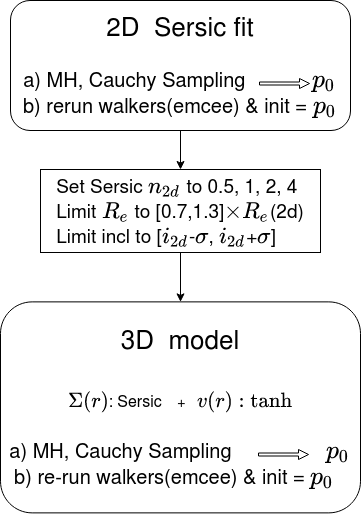}
 \caption{Schematic description of our 3D methodology using a 2-step process. First, we determine the morphological parameters by fitting a 2D \sersic\ model for the flux profile $\Sigma(r)$ on a \OII\ image from MUSE. Then, we fit the MUSE \OII\ data with 3D model with the \sersic\ $n$ fixed to 0.5, 1, 2 or 4 based on $n_{\rm 2d}$ and priors on the half-light radius $R_{\rm e}$ and inclination $i$ from the results of the 2D analysis. We run the MCMC algorithm in two steps, first with Cauchy sampling which reaches convergence more efficiently, then with \textsc{emcee} which better sample the posteriors.}
 \label{fig:method3d}
 \end{figure}

Furthermore, as illustrated in Fig.~\ref{fig:method3d}, we use the following two-step process. We first determine the morphological parameters by fitting a 2D \sersic\ model to the collapsed \OII\ cube (i.e. a two-dimensional flux map), keeping the inclination $i$ free. This yields estimates of the morphological parameters, 
 namely inclination, size and \sersic{} index ($i_{\rm 2d}$, $R_{e,\rm 2d}$, $n_{\rm 2d}$). Then, we fit a 3D model with the \sersic\ $n$ fixed to 0.5, 1, 2 or 4 based on $n_{\rm 2d}$, and tophat priors on the half-light radius $R_{\rm e}$ and inclination $i$ from the results of the 2D analysis (\Fig{fig:method3d}).
 For the kinematics, we choose a  $\tanh(r/r_t)$ rotation curve $v(r)$ after finding that it gives more robust measurements of $V_{\rm max}$ than a $\arctan$ profile. 
 Finally, from the morphological and kinematics information, namely $\Sigma(r)$ and $v(r)$ we can compute $j$ using \Eq{eq:AM:2d}. The total sAM is then calculated from the $j(<r)$ profile at 10~$\Re$~\footnote{Others uses $j(<\Re)$ \citep{CorteseL_16a} or even $j(<R_{\rm max})$ where $R_{\rm max}$ is the last radius where $\Sigma_\star(r)$ can be measured \citep[e.g.,][]{PostiL_18b,MarascoA_19a,ManceraPinaP_21a}, which is prone to biases.}.

For the velocity dispersion, we evaluate the intrinsic (deconvolved) velocity dispersion profile $\sigma(r)$  at $2\Re$, namely   $\sigma_{\rm t}\equiv\sigma(2\Re)$. This is a more reliable estimator of the
dispersion in the outer regions from our parametrization discussed earlier. Indeed, the fitted parameter, $\sigma_0$, can approach zero in some galaxies when the kinematic dispersions ($\sigma_{1,2}$) is sufficient to describe the data.  In the remainder of the paper, we   use $\sigma(2\Re)$ as the total velocity dispersion $\sigma_{\rm t}$ of SFGs. 

 In summary, our 3D method has several advantages and avoids the shortcomings mentioned in the previous section. In particular,
 it includes all of the information available by combining the many spaxels with very low signal-to-noise ratio (S/N), down to $\sim$0.1 in the outer regions, which are inaccessible with 2D velocity maps.
 Second,  it avoids the assumptions of the \Eq{eq:RF12} formula, namely of flat rotation curves, which can lead to biases in the  low-mass regime.
 Third, it takes into account the effect of the PSF convolution. And fourth, it is self-consistent and does not rely on external data to break the $i-V_{\rm max}$ degeneracy.
It does, however, assume that SFGs have regular velocity fields.  Our  modeling is similar to   the {\textsc DYSMAL} code used in \citet{BoucheN_07b,CresciG_09a,DaviesR_11a,GenzelR_17a,GenzelR_20a}, but \galpak\
adjusts the model parameters directly to the raw 3D cube while {\textsc DYSMAL} fits the 1D flux, velocity and dispersion maps generated from a 3D model.
 
\section{Methodology validation with the TNG50 simulation}
\label{section:Illustris}

As our source of simulated galaxies we employ the TNG50 simulation \citep{NelsonD_19a,PillepichA_19a} from the IllustrisTNG project \citep{SpringelV_17a,PillepichA_17a,NelsonD_17a,MarinacciF_17a,NaimanJ_17a}. TNG50 is a cosmological hydrodynamical simulation following a volume of $(51.7\Mpc)^3$ at a baryonic (DM) mass resolution of $8.5\times10^4\Msun$ ($4.5\times10^5\Msun$) and a spatial resolution of  $\sim300\pc$. The simulation is evolved with the \AREPO{ }code \citep{SpringelV_10a}, { and includes physical models for radiative cooling, star-formation, stellar evolution, black hole growth, as well as feedback channels in the form of galactic winds and several modes of AGN feedback. We refer the reader to \citet{VogelsbergerM_13a}, \citet{WeinbergerR_16a} and \citet{PillepichA_16a} for full details on these models and only describe here the one most directly relevant to this work, namely the star-formation model. Star-formation in TNG50 is based on \citet{SpringelV_03a} and prescribes the star-formation rate as a function of volumetric gas density with a threshold of $n_{{\rm H},th}=0.13\cm^{-3}$ and a gas consumption timescale of $2.2\sqrt{n_{{\rm H},th}/n_{{\rm H}}}\Gyr$}. The TNG simulations reproduce rather well several basic properties of galaxies that are relevant to the topic of this paper, such as galaxy stellar mass functions \citep{PillepichA_17a} and size-mass relations \citep{GenelS_17a} at various redshifts.

The angular momentum of galaxies was studied by \citet{GenelS_15a} in the precursor Illustris simulation \citep{VogelsbergerM_14a,VogelsbergerM_14b,GenelS_14a} and was found to match $z=0$ observations within the uncertainties.

\subsection{Sample selection}
\label{section:tng50:selection}

\begin{figure*}
\centering
\includegraphics[width=9cm]{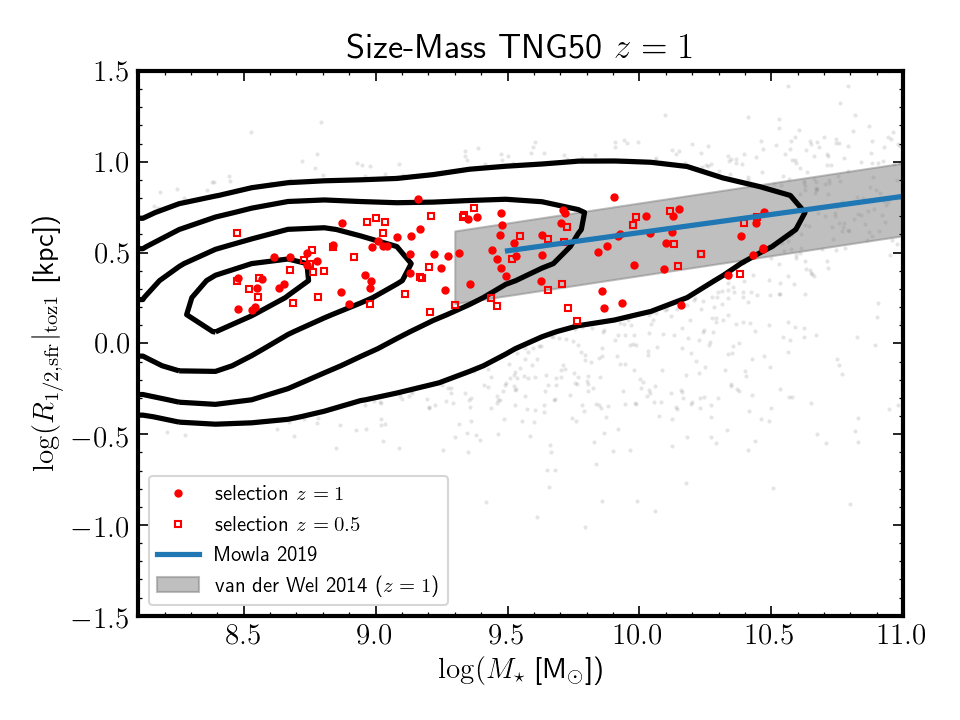}
\includegraphics[width=9cm]{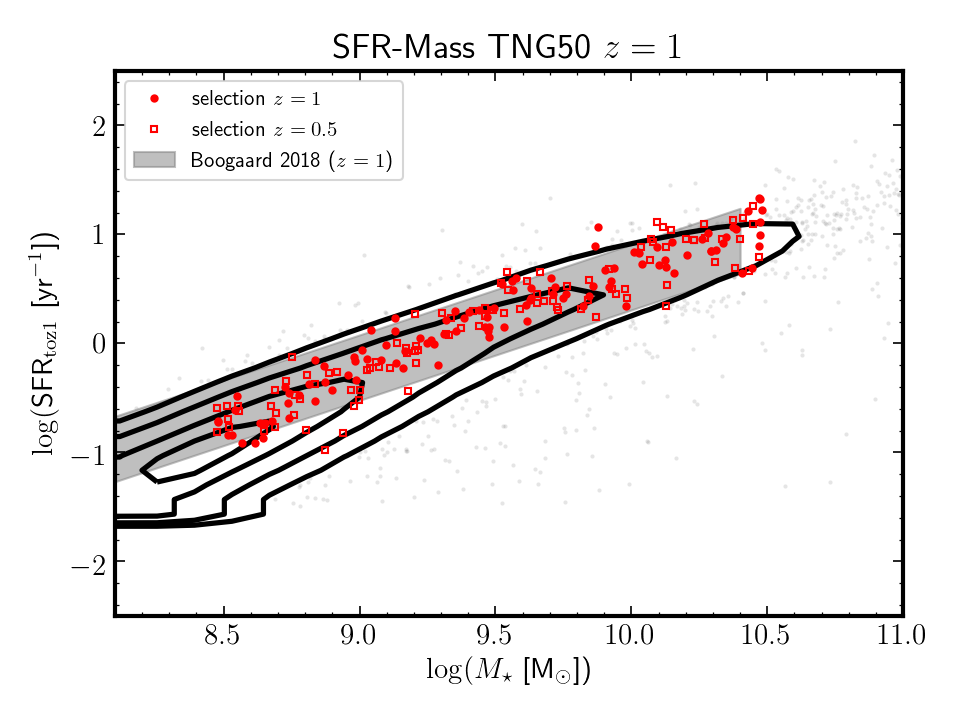}
\caption{Sizes and SFRs of the \Ntng{} galaxies selected from the overall TNG50 galaxy population. {\it Left:} The size-mass relation for the selected TNG50 for the mocks where the sizes are half-SFR sizes ({ calculated in face-on projection and without any dust attenuation}). The open (solid) squares represent the $z=0.5$ ($z=1$) galaxies where the sizes for the $z=0.5$ galaxies are adjusted to $z=1$ according to the $(1+z)^{-0.5}$ evolution found by \citet{vanderWelA_14a}. The background gray points and contours represent the size-mass distribution for the $z=1$ TNG50 galaxy population. The solid line (gray band) represents the size-mass relation for late types from \citet{MowlaL_19a} \citep{vanderWelA_14a}, respectively.
{\it Right:} The SFR-mass or ``main-sequence'' relation for the selected TNG50 galaxies (red dots, where SFRs for the $z=0.5$ galaxies are adjusted to $z=1$ according to the $(1+z)^{2.2}$ evolution found by \citet{WhitakerK_14a}). The background gray points and contours represent the SFR-mass distribution for the $z=1$ TNG50 galaxy population. For comparison, we show with the gray band the $z=1$ main-sequence from the MUSE UDF analysis of   \citet{BoogaardL_18a}.
}
\label{fig:tng:selection}
\end{figure*}


We select 100 SFGs from TNG50 at each of the  $z=0.5$ and $z=1.0$ snapshots with stellar masses in the range $10^{8.5}\Msun<M_{\star}<10^{10.5}\Msun$. We selected SFGs with specific SFR within $\pm2\sigma$ from the median sSFR, 
excluding galaxies with sSFR$<10^{-11}\yr^{-1}$ in order to mimic the observational selection on SFGs with \OII\ emission. 
From this sample, we randomly select the final sample to have a uniform distribution in $\log M_{\star}$ and limited the sample to  the range of sizes matching the UDF sample, namely from 1.5 to 6.5 kpc (see \S~\ref{section:UDF}).
This resulted in a sample of \Ntng{} SFGS.

\Fig{fig:tng:selection} shows the location of  the sample of \Ntng{} SFGs in the size--mass and sSFR--mass scaling relations. In both panels, the gray points and contours represent the distributions for the $z=1$ TNG50 SFGs and the red symbols represent the \Ntng{} selected SFGs at $z=0.5$ and $z=1.0$.
On the left panel, the gray band represents the size--mass relation from \citet{vanderWelA_14a} and on the right panel, the gray band represents the main-sequence for UDF SFGs \citep{BoogaardL_18a}. 
In the left (right) panel, the $z=0.5$  galaxies are rescaled to $z=1$ according to the size (SFR) evolution  found by \citet{vanderWelA_14a} \citep{WhitakerK_14a}, respectively.
 This figure shows that we selected SFGs with properties representative to the full underlying population around the size and SFR scaling relations for  SFGs.

\begin{figure*}
\centering
\includegraphics[width=17cm]{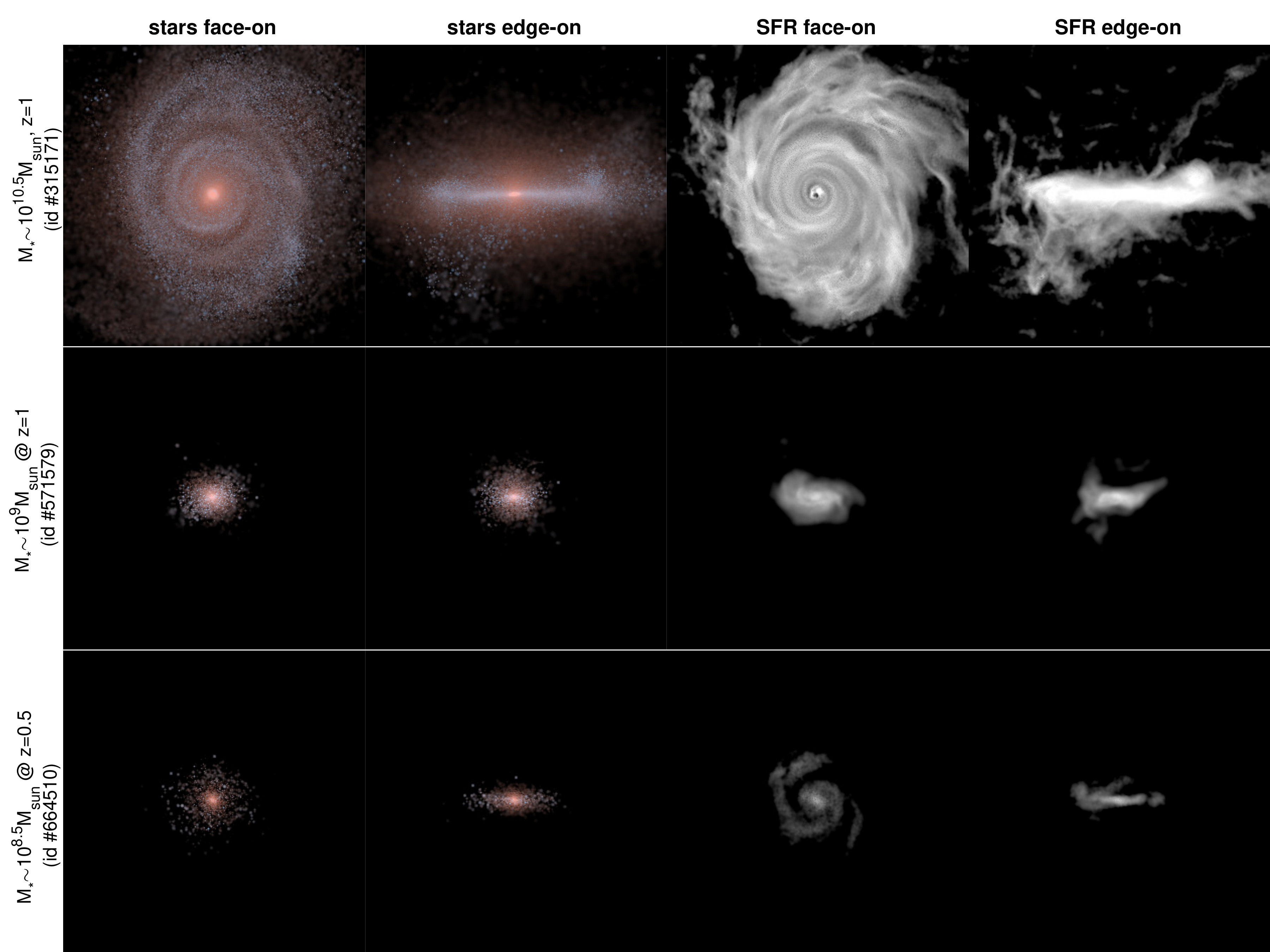}
\caption{Visual representations of three example TNG50 star-forming galaxies from the set analyzed in this work. The two columns on the left represent stellar light (dust-free color composite of the (SDSS)r-g-(Johnson)B bands) and the two on the right the SFR in the gas phase. Each panel is $30\kpc$ on a side.}
\label{fig:galimages}
\end{figure*}

 Finally, \Fig{fig:galimages} shows high-resolution images of three representative star-forming galaxies
 where the two left columns show the stellar light (dust-free color composite of the (SDSS)r-g-(Johnson)B bands) 
 and the two right columns show the SFR in the gas phase, face-on and edge-on.


\subsection{Simulated MUSE cubes}
\label{section:tng:mocks}

Each of the \Ntng\ simulated SFGs  is processed in the following manner to generate several mock  MUSE mini-cubes, for a total number of mock cubes of \Nmock. First, a cube of two spatial dimensions with bounds $[-N_r/2\Delta_r,N_r/2\Delta_r]$ and one velocity dimension with bounds $[-N_v/2\Delta_v,N_v/2\Delta_v]$ is generated with $N_r=30$ pixels in each spatial dimension and $N_v=40$ pixels in the velocity dimension. For $z=1$ ($z=0.5$), we use voxels of size $\Delta_r=1.65\kpc$ (1.25\kpc) and $\Delta_v=50~\kms$ ($65~\kms$), respectively, which corresponds to the MUSE instrument specifications of $0.2\arcsec$ and 1.25\AA{ }at a rest-frame wavelength of $\lambda_{\rm rest}=$3728\AA{ }\citep{BaconR_17a}. Second, the center of this cube is placed at the position of the most bound resolution element of the simulated galaxy, which generally is very close to the center of rotation, and at the velocity of the galaxy center-of-mass. Third, the cube is rotated such that the plane defined by the two spatial dimensions is perpendicular to the eigenvector of the moment of inertia tensor of the galaxy that has the smallest eigenvalue, such that the galaxy appears ``face-on''.

For each galaxy, we apply further rotations to generate mock cubes for 5 inclinations ($i=15,30,45,60$ and $75^\circ$), each at a random azimuthal orientation $\phi$~\footnote{In addition, 9 galaxies were tested with 5 different azimuthal orientations for each inclination.}. The SFR in each voxel of the cube is then calculated by summing up the contributions from all simulation gas cells using a Gaussian kernel representing the Point-Spread-Function (PSF) and Line-Spread-Function (LSF) appropriate for non-AO MUSE observations. In particular, for $z=1$ ($z=0.5$) the PSF has a Moffat profile with $\beta=2.5$ and a ${\rm FWHM_r}$ of $0.625$\arcsec{ }($0.675$\arcsec) \citep[][]{BaconR_17a}, and the LSF has a Gaussian profile with ${\rm FWHM_v}=2.55$\AA{ }(2.75\AA), respectively. The PSF and LSF FWHMs correspond to $5.15\kpc$ ($4.25\kpc$) and to 
$100\kms$ ($143\kms$), for  $z=1$ ($z=0.5$), respectively.

Finally, we add noise to the mock MUSE cubes in order to match the range of S/N (spaxel$^{-1}$)  of our MUSE observations, which have S/N$_{\rm max}$ of a few to $\sim$40 (see \S~\ref{section:UDF}). For galaxies with $\log M_\star>9$ ($\leq9$), we set the noise to 0.001 (0.0002) SFR~yr$^{-1}$,  respectively.


\subsection{Reference measurements}

 In TNG50, we measure the following quantities: the sAM, the maximum rotation velocity $V_{\rm max}$ and velocity dispersion $\sigma_{\rm t}$, as follows.

{\it Specific angular-momentum.}
The SFR-weighted $\js(r)$ is calculated using \Eq{eq:AM:3d} with $\rho(\vec{x})$ as the star-forming gas density, which corresponds to a \OII\ weighting,
 using the 3D positions, velocities and densities of all SF gas at a given radius. We then use the cumulative profile $\js(<r)$ to then define the total sAM from  max$(\js(<r))$.

{\it Velocity dispersion.} The gas velocity dispersion $\sigma_{\rm t}$ is defined as the 1D line-of-sight dispersion assuming isotropy, namely it is equal to $\sigma_{3D}/\sqrt{3}$. Here, $\sigma_{\rm 3D}$ is the average (SFR-weighted) of the internal velocity dispersion of star-forming gas inside kpc-sized patches at galacto-centric radii of 1.5--2.5$\Re$, a method similar to \citet{PillepichA_19a}.
In order to account for thermal broadening for $T=10^4$ K gas, we incorporate a $\sigma_{\rm th}\approx15.7~\kms$ component into $\sigma_{\rm 3D}$ by adding it in quadrature\footnote{This approach follows in spirit the calculations in Appendix A of \citet{PillepichA_19a}. However, it is important to note that \citet{PillepichA_19a}, in method (ii) of their appendix, have inadvertently added $\sigma_{\rm th}\approx15.7~\kms$ to their line-of-sight velocity dispersions, while for a $T=10^4$ K ideal monoatomic Hydrogen gas that value corresponds to the 3D velocity dispersion, as implemented in this work.}.
This definition of line-of-sight SFR-weighted velocity dispersion at $\sim2\Re$ has the advantage that it closely mimics our observations   $\sigma_{\rm t}(2\Re)$ (see \S~\ref{section:methodo:3D}).
 
{\it Rotational velocity.}
The rotation velocity $V$ is calculated as the SFR-weighted mean of the azimuthal velocities of the same patches used for the calculation of the velocity dispersion.


\subsection{Robustness of $\js$(3D) from TNG50}
\label{section:tng50:results}

\begin{figure*}
\centering
\includegraphics[width=\textwidth]{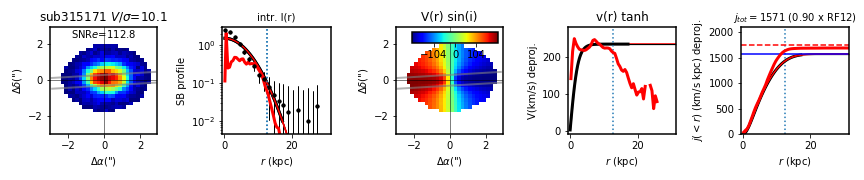}
\includegraphics[width=\textwidth]{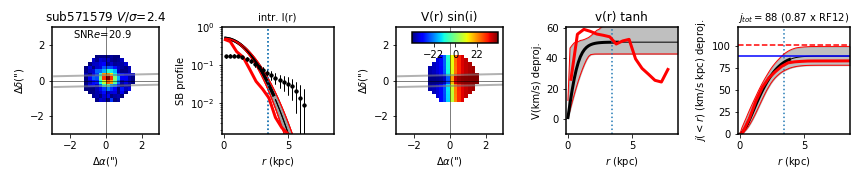}
\includegraphics[width=\textwidth]{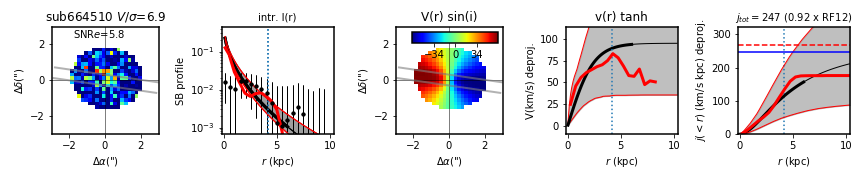}
\caption{Example of \galpak\ fits for  three SFGs from  TNG50  shown in \Fig{fig:galimages}. 
The panels show the flux map, the  flux profile $\Sigma(r)$, the intrinsic velocity field, the deprojected velocity profile and the deprojected angular momentum profile, from left to right respectively. The 1D profiles are computed along the pseudo-slit shown in panel 1 and 3 by the gray lines.
In panel 2, the SB profile for a single orientation is shown (black points).
In panel 2, 4 and 5, the solid gray curves show the model fits for the various inclinations and orientations. The red solid lines represent the true profiles  determined directly from the TNG50 data. The vertical dotted lines represent twice the  half-light radius ($R_{1/2,\rm SFR}$).
The horizontal blue lines show the total modeled angular momentum $j_{\rm 3D}$, while the horizontal red lines show the total $j$ from RF12 (Eq.~\ref{eq:RF12}), which can lead to over-estimation of the total angular momentum by 10--20\%. 
}
\label{fig:jprof:example}
\end{figure*}

\begin{figure*}
\centering
\includegraphics[width=0.48\textwidth,height=8.5cm]{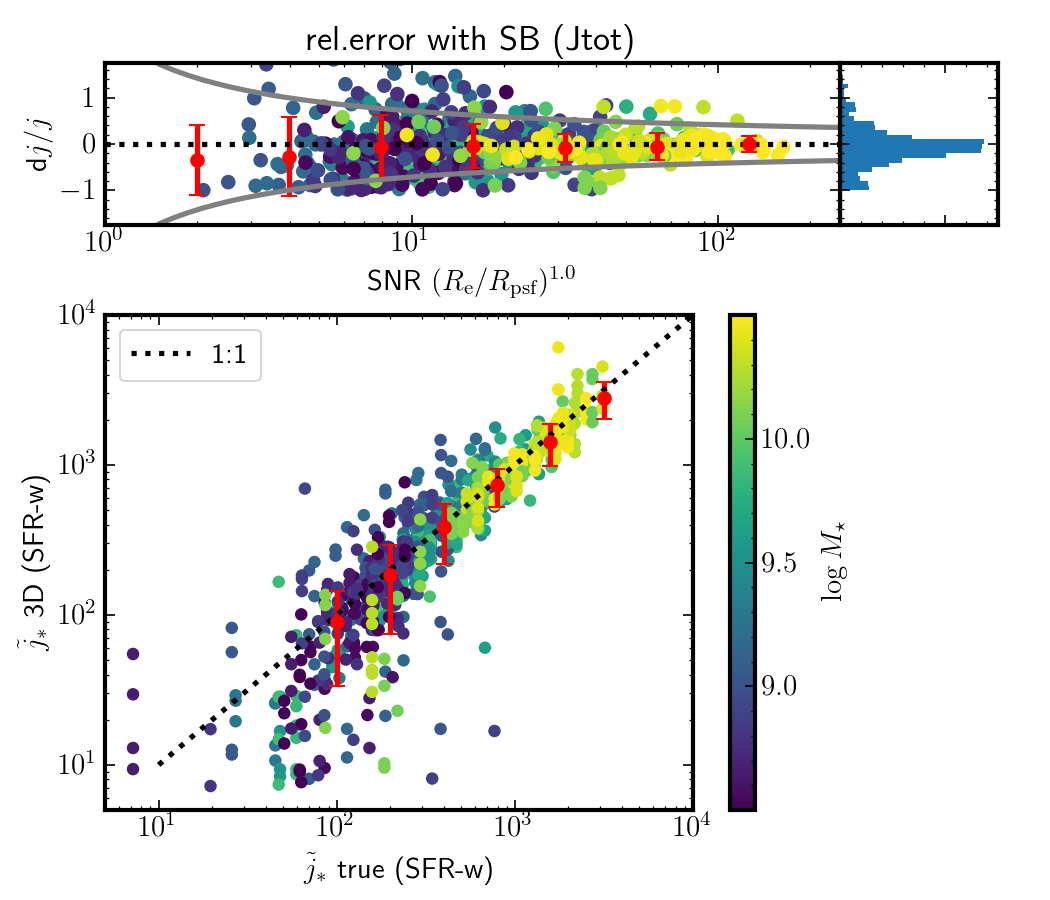}
\includegraphics[width=0.50\textwidth]{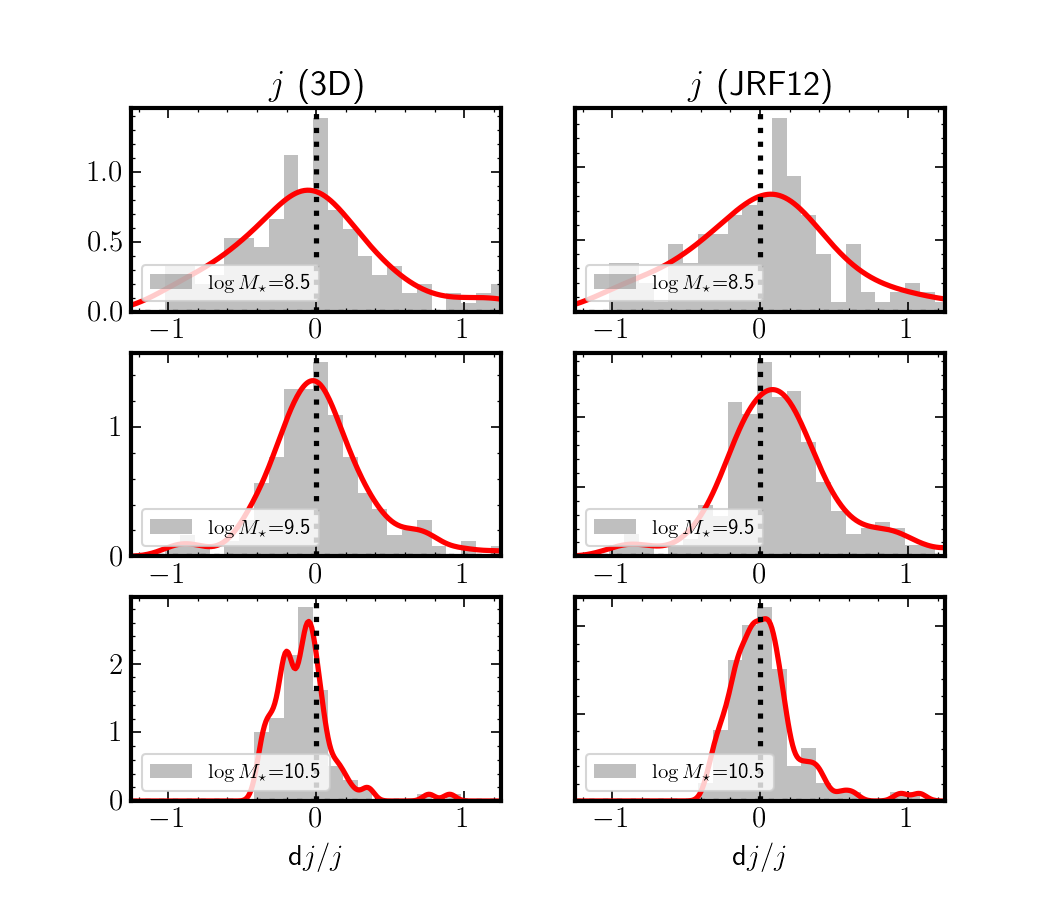}
\caption{Reliability of our sAM estimator. {\bf Left:} The bottom panel shows the total SFR-weighted sAM, $\js$ (in $\kms~\kpc$), as a function of the true angular momentum (SFR-weighted) for the \Nmock{} mock galaxies. The top panel shows the relative error ($\js-j_{\rm true})/j_{\rm true}$ as a function of the effective S/N (see text) along with the histogram of ${\rm d}\js/\js$.
The points are colored according to the stellar mass in both panels.
The red circles  represent the median ${\rm d}\js/\js$ whose errorbar represent $1\sigma$.
 \label{fig:dJ}
{\bf Right:} The relative error (${\rm d}j/j\equiv\js-j_{\rm true})/j_{\rm true}$ for each mass bin of 1~dex centered on $\log M_\star=8.5, 9.5, 10.5$, from top to bottom. The histograms are weighted by the errors.
The left (right) panels show the distributions of  ${\rm d}\js/\js$, in three mass bins, resulting from the 3D method (the \Eq{eq:RF12} approximation), respectively. The red solid line represents the continuous density estimation of the distributions using a Gaussian kernel with bandwidth of $\sim$0.2. This shows that, for massive galaxies ($M_\star>10^{10}\Msun$), the  \Eq{eq:RF12} approximation is less biased, while for galaxies with $M_\star<10^{10}\Msun$, the 3D method is less biased.
 }
\label{fig:dJ:hist}
\end{figure*}



We now apply the methodology described in \S~\ref{section:methodo:3D}
to the sample of \Nmock{} mock cubes of \Ntng{} simulated galaxies  in order to test  the performances of \galpak{}  in recovering the morphological (sizes, inclination) and kinematic parameters of realistic galaxies. We find that the inclination is recovered with a relative error of $\pm12^\circ$ provided that $i\geq30^\circ$, as found in \citet{ContiniT_16a}  \citep[see also][]{BoucheN_15a}. Consequently, we   restrict the analysis to mock cubes with $i\geq30^\circ$.

\Fig{fig:jprof:example} shows the sAM profile $\js(<r)$ found when applying Eq.~\ref{eq:AM:2d} and the 3D methodology described in \S~\ref{section:methodo:3D} on the massive galaxy shown at the top of \Fig{fig:galimages}. 
The 5 panels show the $\Sigma_{\rm sfr}$ map, the SB profile, the intrinsic velocity field, the intrinsic velocity profile, and the intrinsic $\js$ profile, respectively. The velocity and SB profile are computed on the pseudo-slit, represented by the two horizontal gray lines.
The light gray solid lines represent the profiles recovered from our algorithm for each inclination ($i\geq30^\circ$), each performed on 5 orientations in this case.
The red solid lines in columns 2, 3 and 4 represent the $\Sigma_{\rm sfr} $, $v(r)$ and $\js$   
profiles  determined directly from the TNG50 data. 
In the right most panel, the blue horizontal tick marks show our estimates  $\js({\rm 3D})$, while the red horizontal tick marks show $\js$ from RF12 (Eq.~\ref{eq:RF12}).  Depending on the steepness of the $v(r)$ profile, the RF12 approximation can lead to an over-estimation of the total angular momentum by 15--25\%\ for the reasons described in \S~\ref{section:methodo}. \Fig{fig:jprof:example} suggests that our 3D method is able to recover the intrinsic $\js(<r)$ profile with less bias, even when the rotation curve deviates from our assumed   parametrization $v(r)=\tanh$.
Next, we investigate the difference between the estimated and true $\js$ as a function of   S/N.


Figure~\ref{fig:dJ}(left) shows the measured total angular momentum, $\js$(3D), as a function of the true angular momentum (SFR-weighted) as measured directly from the simulation output (bottom subpanel), and their relative difference, ${\rm d}j/j\equiv (\js({\rm 3D})-j_{\rm true})/j_{\rm true}$ (top subpanel).  The points are colored according to $M_\star$.
The red circles represent the median of ${\rm d}j/j$ whose errorbars represent the $1\sigma$ standard deviation.

The histogram in the top panel of Fig.~\ref{fig:dJ}(left) shows the distribution of the relative errors ${\rm d}j/j$. This histogram shows that globally our method does recover the intrinsic angular momentum with no apparent bias.  We can quantify this bias further as a function of galaxy mass in Fig.~\ref{fig:dJ:hist}(right).
Figure~\ref{fig:dJ:hist}(right) shows, for 1-dex bins in $\log M_\star$ (at 8.5, 9.5 and ${10.5}$), from top to bottom), the distributions of  ${\rm d}j/j$ resulting from  our 3D method (left column) or the RF12 method (right column). The red solid line represents the continuous density estimation of the distributions using a Gaussian kernel with bandwidth of 0.2. Quantitatively, the mean relative error ${\rm d}j/j$ for our 3D methodology is $<5$\%, while for the intermediate and lower mass bins, the RF12 approximation (Eq.~\ref{eq:RF12}) can lead to a significant bias of $\approx10$-15\%\ for the reasons discussed in \S~\ref{section:methodo}.  For the most massive galaxies with $M_\star>10^{10}\Msun$, and with $\log \js>3$, our method underestimates $\js$ compared to RF12, but it is important to note that our sample in the UDF (\S~\ref{section:UDF}) has only a handful of SFGs in this regime.

The width of the distributions shown in \Fig{fig:dJ:hist} is consistent with the errors obtained with our methodology. Indeed, the width of  the ratios of the errors to ${\rm d}j$ is consistent with a (Cauchy)  distribution~\footnote{The few outliers are producing a distribution with a heavier tail than a Normal distribution.}  of width unity. This implies that the errors obtained from our methodology are broadly representative of the true error ${\rm d}j$ on the measurement.

\begin{figure}
 \centering
 \includegraphics[width=9cm]{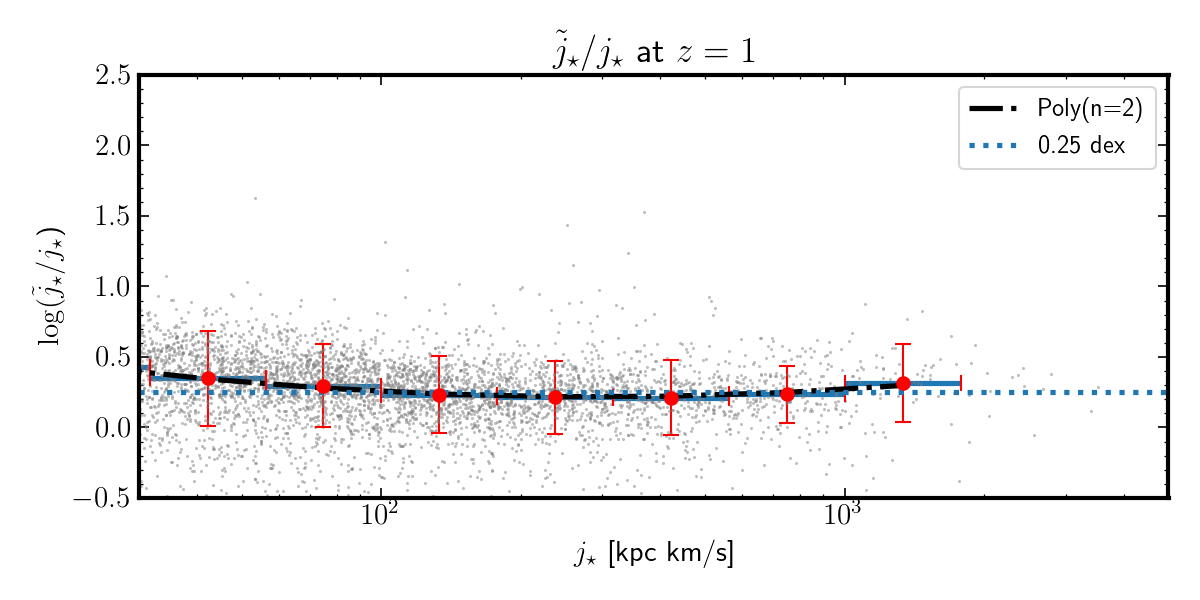}
 \caption{Relation between the stellar sAM $j_\star$ and $\js$ for $z=1$ TNG50 SFGs. The solid points with errorbars represent the median difference between the two as a function of the stellar angular momentum, where the error bars represent the standard deviation.
 Thus, the SFR-weighted sAM, $\js$, is higher by $\approx0.25\dex$ (dotted line) compared to $j_\star$ over a wide range of sAM from 50 to 2000 $\kms$~kpc. 
 A second order polynomial fit is shown with the dot-dashed black line.
 }
 \label{fig:Jcomparison}
 \end{figure}

We return now to the top panel of Figure~\ref{fig:dJ} where the relative difference ${\rm d}j/j$ is plotted as a function of the effective S/N, S/N$_{\rm eff}$, following the discussion in \citet{BoucheN_15a}.
This effective S/N accounts for the fact that
 surface brightness (SB) alone is not sufficient to determine the accuracy in the fitted parameters, given that the
compactness of the galaxy with respect to the PSF plays also an important role.
As discussed in \citet{RefregierA_12a},
in the presence of a PSF convolution,  the relative errors on the major-axis $a$ scales as
\begin{eqnarray}
\frac{\sigma(a)}{a}&\propto& A_{\rm o}^{-1}(1+R_{\rm PSF}^2/\Rhalf^2),\label{eq:errorp}
\end{eqnarray}
where $A_{\rm o}$ is the observed central surface brightness in the central voxel,  $R_{\rm PSF}$ is  the radius of the PSF ($R_{\rm PSF}\equiv$ FWHM/2) and $\Rhalf$ the {intrinsic} half-light radius.  
After performing a Taylor
expansion around
$R_{\rm PSF}/\Rhalf \sim (1-x)$ with $x\equiv (\Rhalf-R_{\rm PSF})/\Rhalf$ and $|x|<<1$,
one finds that  Eq.~\ref{eq:errorp}    becomes in the regime where $R_{\rm PSF}/\Rhalf$ is $\simeq1.0$ \citep[see][]{BoucheN_15a}:
\begin{eqnarray}
\frac{\sigma(a)}{a}&\propto&  \left(\frac{\Rhalf}{R_{\rm PSF}}A_{\rm o}\right)^{-1}\nonumber\\
&\propto& \left(\frac{\Rhalf}{R_{\rm PSF}} {\rm S/N}_{\rm max}\right)^{-1}.
\label{eq:refregier}
\end{eqnarray}
The last step follows from the simple argument that  a given exposure time sets the surface brightness limit per voxel, and hence the maximum  S/N (in the central voxel or spaxel), S/N$_{\rm max}$, is directly proportional to  $A_{\rm o}$.

Considering the sAM is essentially determined by the product of $\Rhalf$ and $V_{\rm max}$, one can thus expect that the accuracy of $j$ follows \Eq{eq:refregier}, which is represented by the gray curve in the top panel of  \Fig{fig:dJ}.

Finally, we end this section by quantifying the potential difference between the SFR-weighted $\js$ and the stellar mass-weighted $j_\star$ because in observations of high-redshift galaxies, as in \S~\ref{section:UDF}, one is often limited to the SFR-weighted $\js$. 
Using TNG50  $z=1$ SFGs in the parent sample, Figure~\ref{fig:Jcomparison} shows the relation between $\js$ and $j_\star$.
While this figure shows that  the relation becomes non linear in the low mass regime, with  $\log j_\star<1.5$,
\citep[as found by][in the FIRE simulations]{ElBadryK_18a}, it also shows that, on average,
$\log \js$ is an estimate of $\log j_\star+0.25$ dex over a wide range of sAM from 50 to 2000 $\kms$~kpc, which is the relevant range for our study.



\section{Application to UDF galaxies}
\label{section:UDF}

\subsection{Sample}

We exploit the 3'$\times$3' 
observations taken with the MUSE instrument over the {\it Hubble}
Ultra Deep Field \citep[HUDF;][]{BaconR_17a}   to investigate the  angular momentum of intermediate redshift galaxies using the  methodology presented in \S~\ref{section:methodo:3D}.  The observations were performed on a mosaic of nine 1'$\times$1' 10~hr pointings  and contain a single 1'$\times$1' 27hr pointing, but we concentrate on the mosaic sample to ensure a homogenous selection. In the MUSE UDF mosaic, there are \Ntotal\  \OII\ emitters \citep{InamiH_17a}.
%
These \OII\ emitters have stellar masses ranging from $10^{7.5}$ to $10^{11}$\Msun, where
the stellar masses  are estimated   \citep[as in ][]{BoogaardL_18a} using the stellar population synthesis (SPS) code FAST \citep{KriekM_09a} assuming a \citet{ChabrierG_03a} initial mass function with an exponentially declining star formation history. 
As discussed in the next section, most \OII\ emitters are rather faint, with only about $\sim$50\%\ of the these  having a S/N  above $\sim\SNRmin$. 
 
\begin{figure*}
 \centering
 \includegraphics[width=0.9\textwidth,height=7cm]{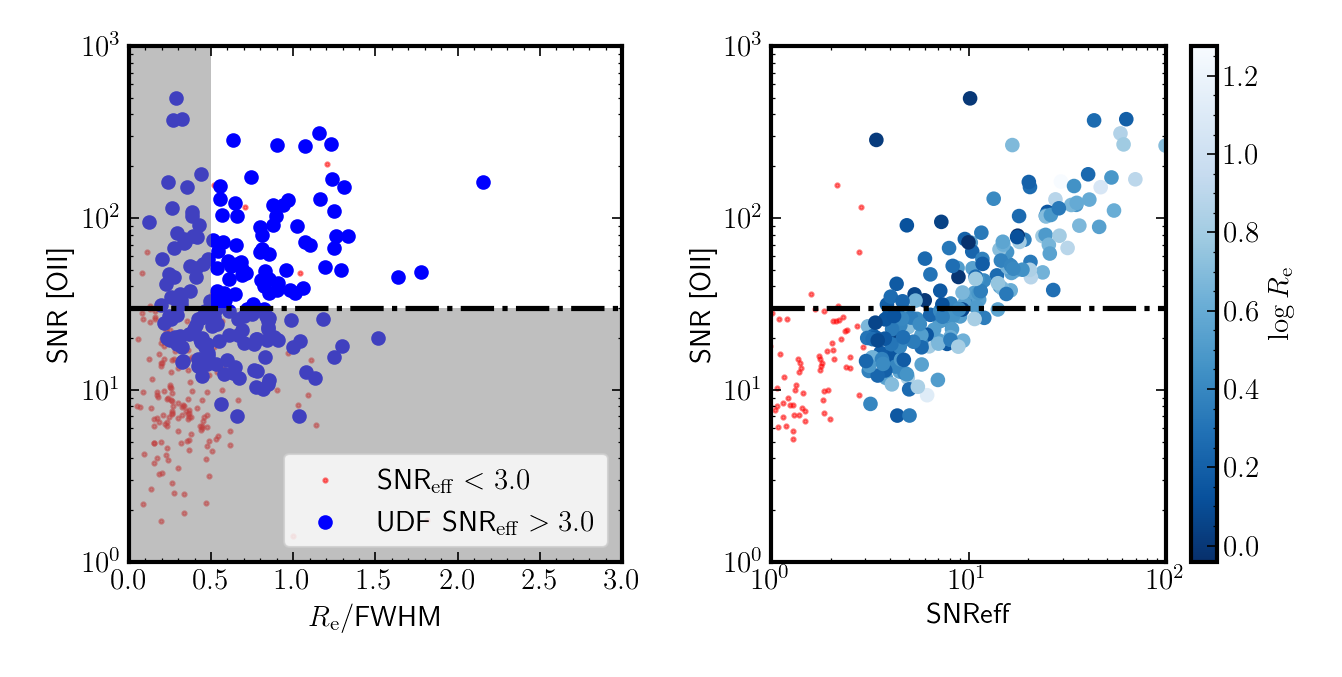}
 \caption{{\bf Left:} Following  \citet{AbrilV_21a}, we show the S/N  of  the  total  \OII{}  flux  as  a  function  of  the resolved ratio ($R_{\rm e,hst}$/PSF FWHM)    for the sample of SFGs. The gray bands represent the selection of resolved galaxies in \citet{AbrilV_21a}, consisting of
S/N$>30$  and $R_{\rm e,hst}/$FWHM>0.5. 
 {\bf Right:} Comparison between the total \OII{} S/N and the effective S/N (\Eq{eq:refregier}), color-coded with $\Rhalf$ .
 In both panels, the red points are galaxies with S/N$_{\rm eff}<3.0$, while the blue points correspond to galasxes with S/N$_{\rm eff}$>3.0.
 \label{fig:udf:selection}}
 \end{figure*}

 {
 Figure~\ref{fig:udf:selection} (left) shows  the S/N  of  the  total  \OII{}  flux  as  a  function  of  the ``resolved size'' ($\Rhalf$ divided by the PSF FWHM)    for the sample of SFGs, where the blue (red) points are SFGs with a S/N$_{\rm eff}>3.0$ ($<3.0$), respectively. For comparison, the selection used in \citet{AbrilV_21a} is shown with the gray bands. The right panel compares the effective S/N and the total S/N, where the points are color-coded as a function of size $\Rhalf$, displaying the common surface-brightness biases. This figure shows that a S/N$_{\rm eff}$  3, as determined in the previous section, opens the possibility to study the kinematics of faint galaxies, enlarging the sample by 30 to 50\%.
 }
 
\begin{figure*}
\centering
\includegraphics[width=0.45\textwidth]{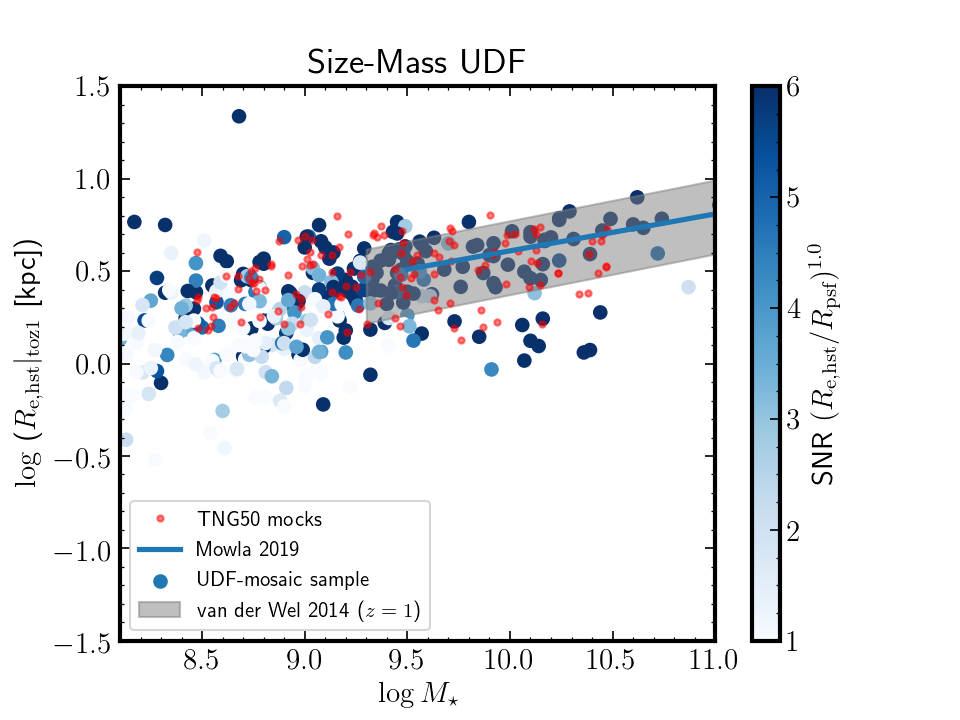}
\includegraphics[width=0.45\textwidth]{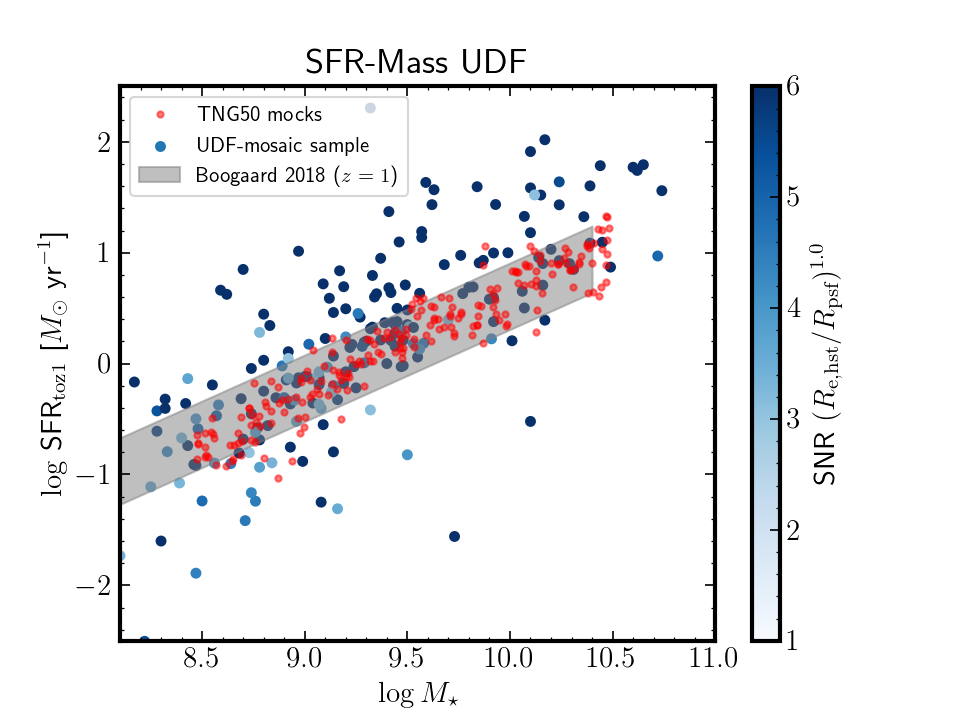}
\caption{Comparison of selection techniques. {\bf Left:} The size-mass relation for the SFGs in the UDF mosaic sample (blue circles).   The sizes are adjusted to $z=1$ according to the $(1+z)^{-0.5}$ evolution found by \citet{vanderWelA_14a}. The solid line (gray band) represent the size-mass relation of \citet{MowlaL_19a} \citep{vanderWelA_14a}.
{\bf Right: } The main-sequence relation for the SFGs in the UDF mosaic sample. The SFRs are adjusted to $z=1$ according to the $(1+z)^{1.74}$ evolution of \citet{BoogaardL_18a} whose $z=1$ main sequence is represented by the gray band.
In both panels, the points are color-coded according to effective S/N. 
For comparison we show the properties of the mock \Ntng{} galaxies  (\S~\ref{section:tng:mocks}) as red circles.
} 
\label{fig:udf:sample}
\end{figure*}

 Figure~\ref{fig:udf:sample} shows the properties of the UDF sample in terms of sizes and SFRs  as in Fig.~\ref{fig:tng:selection}. 
 The sizes are the continuum-based $\Re$ obtained from {HST}/F160W  from \citet{vanderWelA_14a} and the SFRs are derived from SED fitting.
 The left panel shows the sample in the size-mass plane color coded as a function of the effective S/N, where the sizes were adjusted to $z=1$. For comparison, we show the size-mass scaling relation of \citet{vanderWelA_14a} and \citet{MowlaL_19a} down to $M_\star\sim10^8\Msun$.
 The TNG50 mocks used in \S~\ref{section:Illustris} are represented with the red small circles. The right panel shows the main-sequence or the SFR-mass plane where the points are color-code as a function of size. The points are scaled to $z=1$ where we used the main-sequence redshift evolution of \citet{BoogaardL_18a}.
 
\subsection{Size estimates}
\label{section:udf:sizes}

 The first part of the methodology outlined in \ref{section:methodo:3D} is to determine the morphological parameters (size, inclination) by performing two dimensional \sersic\ fits to the \OII\ flux maps. Given that the UDF data set has extensive {\it Hubble} Space Telescope ({HST}) coverage, we first ask whether our \sersic\ fits on \OII{} (from ground-based MUSE observations at a resolution of 0.62") are able to recover the {HST} morphological parameters (at a resolution of 0.15") from their stellar continuum.
 
\begin{figure*}
\centering
 \includegraphics[width=0.45\textwidth]{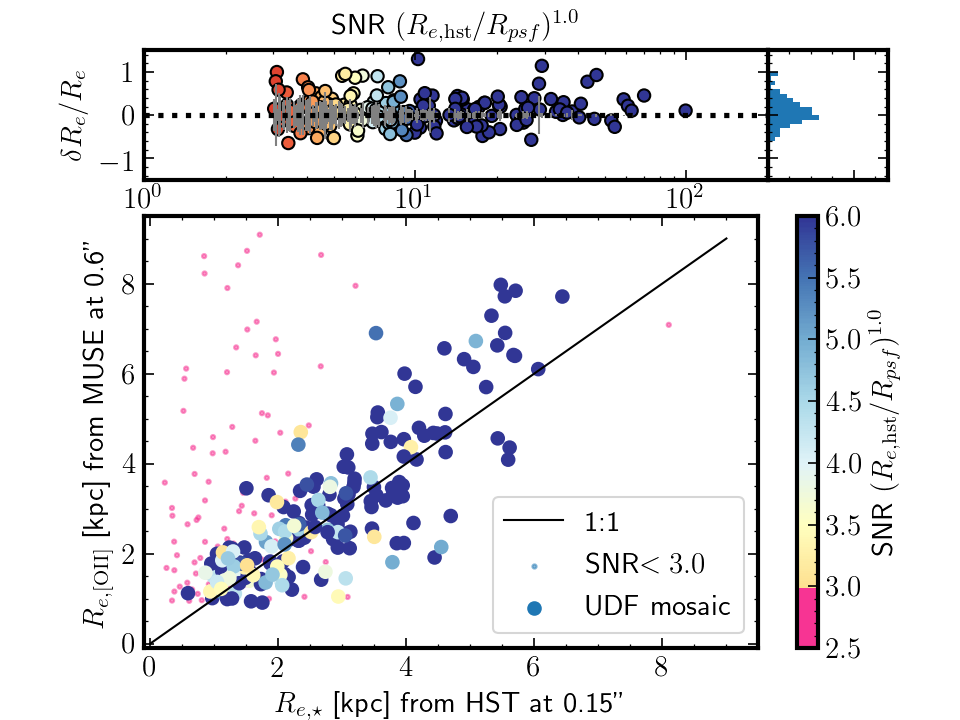}
 \includegraphics[width=0.45\textwidth]{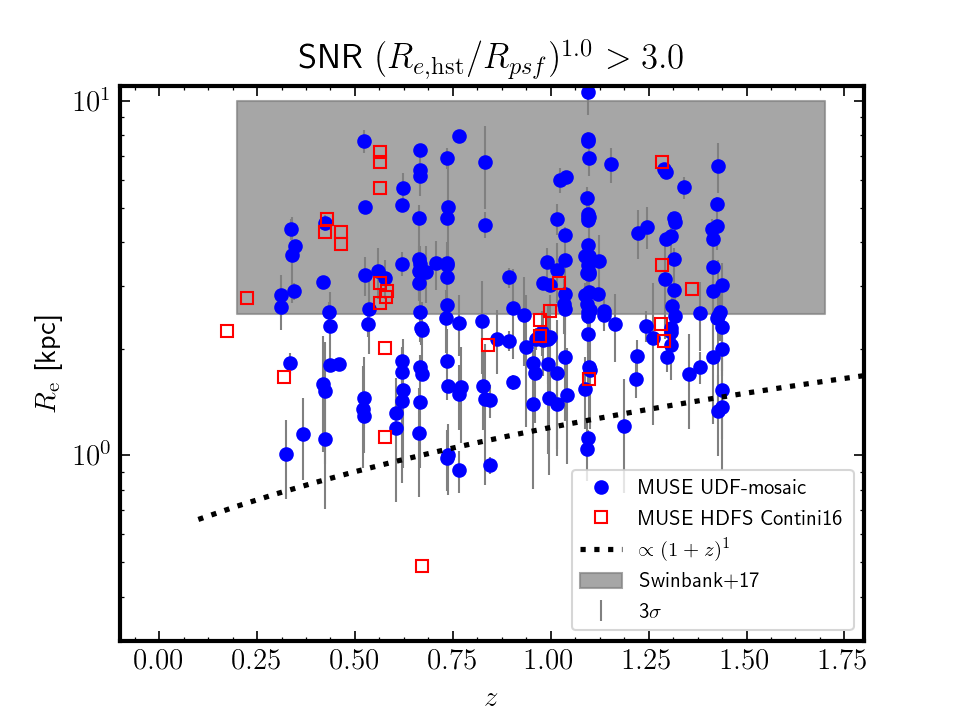}
\caption{{\bf  Left:} Comparison between \OII-based sizes $\Re$ from  MUSE (at FWHM$\approx0.7$" resolution) versus $H$-band continuum $R_{e,\star}$  from {HST}/F160W at $\approx0.15$" resolution, for
the full sample of \Ntotal\ SFGs, where the points are color-coded according to Eq.~\ref{eq:refregier}. The top panel shows the relative error $\delta \Re/\Re$ as a function of the effective S/N (see text) and the side histogram shows its distribution.
The gray errorbars represent the $3\sigma_{\rm t}$ statistical uncertainties.
This shows that our MUSE-\OII{} sizes are of  sufficient accuracy down to $\approx1$~kpc, 
  provided that the effective S/N is $\gtrapprox3$. 
  \label{fig:Re} 
{\bf Right:} The sizes $\Re$ (from HST/F160W) as a function of redshift for \Nsamp\ galaxies in the MUSE UDF-mosaic field whose effective S/N is sufficient.  
 The gray region represents the region probed by  \citet{SwinbankM_17a}. The dashed line represents  $\propto(1+z)^1$. The vertical error bars are $3\sigma$.
 }
 \label{fig:size:redshift}
 \end{figure*}

 Given that the galaxy half-light radius plays a crucial role in determining $j$, we show 
in Fig.~\ref{fig:Re}(a) the intrinsic \OII-based
$\Re$ found from our MUSE data compared to the continuum-based $\Re$ obtained from {HST}/F160W  from \citet{vanderWelA_14a}.
This figure shows that there is a good agreement between the two quantities,   provided that the size-corrected ``effective S/N''  (Eq.~\ref{eq:refregier})  in the brightest voxel, namely  S/N$_{\rm max}\times(\Re/R_{\rm psf})$ where  S/N$_{\rm max}$ is the S/N in the central voxel and $R_{\rm psf}$ is the PSF radius (half-FWHM), is $\gtrapprox\SNRmin$. 
 The S/N level of $\sim\SNRmin$ is our fiducial threshold for the remainder of our analysis. From the \Ntotal\ SFGs in the 9 arcmin$^2$ UDF mosaic, there are \Nsamp\ galaxies meeting this criterion.

This demonstrates that our algorithm is able to  determine $\Re$ with sufficient accuracy down to $\approx1$~kpc. This is beyond the 3~kpc limit of other IFS surveys as  the parameter space between $1<\Re<3$ kpc (and $10^8<M_\star<10^{9.5}\Msun$) has been relatively unexplored so far with any of the large IFS surveys
\citep[e.g.,][]{BurkertA_16a,SwinbankM_17a,HarrisonC_17a}. 
Fig.~\ref{fig:size:redshift} shows the location of half-light radii $\Re$ of our subsample of \Nsamp\ galaxies as a function of redshift where the
parameter space probed by \citet{SwinbankM_17a} is shown as the gray band.  This figure shows that,
compared to \citet{SwinbankM_17a}, our sample includes galaxies with sizes in the range 1 to 3 kpc, previously inaccessible from ground-based observations.

 

\subsection{The angular momentum of SFGs at $0.4<z<1.4$}
\label{section:udf:sAM}

We apply the two-step methodology outlined in \S~\ref{section:methodo:3D} on the MUSE UDF sample of \Ntotal\ \OII{} emitters. 
We find that \Nsamp\ meet our S/N threshold discussed above, namely with effective S/N $\gtrsim$\SNRmin.
The sample is thus made of all \OII\ emitters in the UDF MUSE mosaic with sufficient S/N and is composed of SFGs with \sersic{} profiles that are mostly exponential. Two thirds of the sample has $n=1.0$, while the remainder has $n=0.5$. The distribution of inclinations $i$ is consistent with an unbiased sample, namely $P(\cos(i))\sim b/a$ is flat  for $i>30$ ($b/a<0.86$).
As discussed in \S~\ref{section:methodo:3D}, we compute the angular momentum of the star-forming gas $\js$ by integrating \Eq{eq:AM:2d}  from the flux $\Sigma(r)$ and kinematic $v(r)$ \galpak\ fits.

Fig.~\ref{fig:Jtot}(a) shows the resulting $\js-M_\star$ sequence for the $\Nsamp$ galaxies in our sample. This figure shows the  sAM $\js$ (SFR weighted) as a function of stellar mass $M_\star$ where the points are color-coded
according to $\log(V_{\rm max}/\sigma_{\rm t})$ where $\sigma_{\rm t}$ is the outer velocity dispersion (see \S~\ref{section:methodo:3D})
 measured with \galpak\ \citep[defined as in][]{CresciG_09a}.
The right axis shows the corresponding stellar mass-weighted $j_\star$ from the typical offset found in Fig.~\ref{fig:Jcomparison}.

In Fig.~\ref{fig:Jtot}(a), the dotted-dashed  blue (red) line show the $j_\star-M_\star$ sequence for late (early) type galaxies at $z=0$ from \citet{FallM_18a}, respectively. The dotted line shows the $z=0$ sequence for the sample of disk galaxies of \citet{PostiL_18b} which has a pure power-law sAM $j_\star\propto M_\star^{0.55}$ \citep[see also][]{ManceraPinaP_21a}. The thin black dotted curve represents the selection limit of our observations of $V_{\rm max}^2+\sigma_{\rm t}^2=K^2$ described in Appendix~\ref{sec:appendix:limit}.

In a cosmological framework,
 the $j_\star$ of a galaxy with stellar mass
$M_\star$ that resides in a virialized  halo of mass $\Mvir$ at a given redshift $z$
can be written as \citep[following][]{MoH_98a,NavarroJ_00a,RomanowskyA_12a,ObreschkowD_15a,BurkertA_10a,BurkertA_16a,StevensA_16a,SwinbankM_17a}:
 \begin{eqnarray}
 j_\star&\propto&\lambda f_j f_m^{-2/3} H(z)^{-1/3}\Delta_c(z)^{-1/6}M_\star^{2/3}\nonumber\\
 &\propto& \lambda f_j  f_m^{-2/3}M_\star^{2/3}(1+z)^{-1/2},
 \label{eq:js:formula}
 \end{eqnarray}
 where $\lambda$ is the halo spin parameter,
$f_m \equiv M_\star/\Mh$ is the stellar-to-halo mass ratio, $f_j \equiv j_\star/ j_{\rm h}$
is the $j$ retention factor (or the stellar-to-halo sAM ratio), and the expected $(1+z)^{-1/2}$ redshift evolution follows from $H(z)$, 
the Hubble parameter and $\Delta_c(z)$, the halo over-density relative to
the mean density of the universe \citep{BryanG_98a}.
For constant $f_m$ in massive galaxies,
the $j_\star-M_\star$ relation is expected to 
be $\propto M_{\star}^{2/3}$, while in low-mass galaxies,  $j_\star$ should be $\propto M_\star^{1/3}$  given that abundance matching shows that \citep{MosterB_10a,DuttonA_10a,BehrooziP_13b} $f_m\propto M^{1/2}_\star$. In other words,  the $j_\star-M_\star$ relation should have a slope of $2/3$ ($1/3$) in the high-(low) mass regime, respectively, provided that the retention fraction $f_j$ is constant \citep[see][for detailed discussions]{PostiL_19a}.

Following \citet{RomanowskyA_12a}, Eq.~\ref{eq:js:formula} becomes
 \begin{eqnarray}
 j_\star(M_\star,z)&=&1010\,\hbox{km/s kpc}\frac{\lambda}{0.034}{f_j}{} f_m^{-2/3} M_{\star,10}^{2/3}(1+z)^{-1/2} \label{eq:js:predict}\\
 &=& 715 \,\hbox{km/s kpc}\frac{\lambda}{0.034}{f_j}{} f_m^{-2/3} M_{\star,10}^{2/3}(1+z)_{1}^{-1/2}, \nonumber
 \end{eqnarray}
 where $M_{\star,10}\equiv M_{\star}/10^{10}\Msun$, $\lambda$ the halo spin parameter and $(1+z)_1=(1+z)/2$.
 Eq.~\ref{eq:js:predict} is shown as the solid black line in Fig.~\ref{fig:Jtot}(a), assuming $f_m$ from abundance matching \citep{DuttonA_10a}, $\lambda=0.034$ from DM simulations \citep{MaccioA_07a,BettP_07a,BettP_10a}, and $f_j=1$.

\subsubsection{The $j-M-(V/\sigma)$ relation}
 \label{section:UDF:sAM:VoverSigma}

 In the local universe, a transition between the late to early type $j-M$ sequence has also been observed   \citep{RomanowskyA_12a,FallM_13a,FallM_18a,ObreschkowD_14a,Murugeshan2020}, where the $j-M$ relation is  a function of the bulge-to-total ratio $B/T$. Similarly, \citet{HarrisonC_17a} found, on a sample of 586 SFGs at $z\simeq0.9$, that the $j-M$ relation is a function of \sersic{} index $n$, where sAM decreases with increasing $n$. 
\citet{CorteseL_16a}, using SAMI, showed that   the $j-M$ relation is a function of the stellar spin parameter, $\lambda_R$ \citep[as defined in ][]{EmsellemE_07a}, thus defining a continuous sequence in the sAM-stellar mass plane ($j-M-\lambda_R$).

At intermediate redshifts, \citet{ContiniT_16a}, with only 28 $0.4<z<1.4$ galaxies in the MUSE HDFS,  found that $z\simeq1$ SFGs fill a continuous transition between rotation dominated (with $V/\sigma_{\rm t}>1$)
to dispersion dominated galaxies (with $V/\sigma_{\rm t}<1$) where the two populations followed the late/early type sAM sequence of \cite{FallM_13a}.
This $V/\sigma$ trend was
also present in the high mass sample of \citet{BurkertA_16a}.
Here, with our larger sample of intermediate redshift SFGs from the MUSE UDF mosaic 9~arcmin$^2$ fields, we revisit this question.

To quantify this $j-M-(V/\sigma)$ relation, we show in
Fig.~\ref{fig:Jtot}(b) the normalized AM sequence $\js/f(M_{\star,10})$ as a function of $V/\sigma_{\rm t}$ where the points are color-coded with the redshift $z$. The normalization $f(M)$ removes the mass dependence obtained from the fit (see below).
Fig.\ref{fig:Jtot} clearly shows that the $z\sim1$ AM sequence is actually a 3-dimensional $j-M-(V/\sigma_{\rm t})$ sequence, meaning that the sAM is a strong function of the galaxy dynamical state $V/\sigma_{\rm t}$. This confirms the early results of  \citet{ContiniT_16a} and extends the results of  \citet{BurkertA_16a} (their Fig.2) for high-mass SFGs with $M_{\star}>10^{10}\Msun$ to low mass galaxies with $M_{\star}\sim 10^8\Msun$.

 Following \citet{ObreschkowD_14a} and \citet{FallM_18a} who fit $j_\star(M_\star,{\rm B/T})$ as a 2-dimensional plane in a 3D space, we fit
the AM sequence in Fig.~\ref{fig:Jtot} with a 3-dimensional function in $M_\star$, $V/\sigma_{\rm t}$ and redshift $z$.
For this purpose, we use a quadratic polynomial $f(\log M_\star)$ in $\log M_\star$, a linear  redshift dependence in $\log(1+z)$ and a linear relation with $\log(V/\sigma_{\rm t})$ such that: 
\begin{eqnarray} 
\log\,\js &=&  \alpha\;(\log M_{\star,10})^{2} + \beta\log M_{\star,10} +{k}+{ZP(z)}\;{\hbox{with}}\nonumber\\
ZP(z) &=& a \log(1+z)_{1} + b \label{eq:Jtot:fit}\\
 k&=&\gamma \log[(V/\sigma_{\rm t})_{5}]\nonumber
\end{eqnarray}
 where $\log M_{\star,10}=\log(M_\star/10^{10}\Msun)$, $(V/\sigma_{\rm t})_5$ is $V/\sigma_{\rm t}/5$, and $(1+z)_1=(1+z)/2$. In this parametrization, $a$ captures the redshift evolution $(1+z)^a$ and $b$ is the
 zero-point of the sAM relation at $z=1$ for rotation dominated SFGs with $V/\sigma_{\rm t}=5$ and $M_\star=10^{10}\Msun$.
 
  
  In fitting for the redshift evolution in Eq.~\ref{eq:Jtot:fit}, a key assumption is often made \citep[e.g.,][]{SwinbankM_17a},
  namely that the distribution of $V/\sigma_{\rm t}$ is redshift-independent.  
   However, $V/\sigma_{\rm t}$ is known to be strongly evolving with redshift
   \citep[e.g.,][]{ForsterSchreiberN_06a,GenzelR_08a,EpinatB_12a,KassinS_12a,WisnioskiE_15a,UblerH_19a}.
 This  has two (but related) important consequences. First,  any particular definition of disk galaxies (e.g.,~with a constant cut at $V/\sigma_{\rm t}>1$ or $>2$)   becomes a redshift-dependent selection of the underlying popution.    Indeed, a population with a $V/\sigma_{\rm t}(z)$ distribution evolving with redshift means that a fixed $V/\sigma_{\rm t}$ cut (commonly used in the litterature) will select a changing fraction of the underlying disk population.
 Second,   this evolution should be taken into account if the redshift evolution $(1+z)^a$ captured in Eq.~\ref{eq:Jtot:fit} is to mean the redshift evolution at a fixed dynamical state or for the median of the popultation. 

  Since we aim to compare the sAM sequence $\js-M_\star$ across our wide redshift range 0.4--1.4, the redshift evolution of $V/\sigma_{\rm t}$  can bias the $j-M$ redshift evolution, and cannot be neglected. 
 Thus, 
the $k$ term in Eq.~\ref{eq:Jtot:fit} ought to be
  $ k=\gamma \log[(V/\sigma_{\rm t})_{\rm toz1}/5]$, where $(V/\sigma_{\rm t})_{\rm toz1}$ is $(V/\sigma_{\rm t})$ corrected to $z=1$. 
   This ensures that we are describing the evolution of the $\js-M_\star-(V/\sigma_{\rm t})$ sequence  without being impacted from the redshift evolution of $V/\sigma_{\rm t}$. 
 
To estimate $(V/\sigma_{\rm t})_{\rm toz1}$, we use the redshift evolution of  $V/\sigma_{\rm t}$, discussed extensively in \citet{UblerH_19a} and \citet{PillepichA_19a},  which
 comes  primarily from the  evolution of $\sigma_{\rm t}(z)$  
   given the negligible evolution of the Tully-Fisher relation   \citep[e.g.,][]{DiTeodoroE_16a,UblerH_17a,TileyA_19a}.
 It   can be parametrized as
\begin{equation}
\left(\frac{V}{\sigma_{\rm t}}\right)(z)\propto(1+z)^{q}\label{eq:VoverS:evolution}.
\end{equation}
Ignoring this effect will increase bias the redshift evolution of the $j-M$ relation, specifically the $a$ coefficient in Eq.~\ref{eq:Jtot:fit} will be biased by approximately $q\times\gamma$.

We find that the parameter $q$ in \Eq{eq:VoverS:evolution} can be found by minimizing the $V/\sigma_{\rm t}$ distribution over the full redshift range (as discussed in    Appendix~\ref{sec:appendix:VSigma}).
Fig~\ref{fig:appendix:VoverSigma}(right) shows that  $q\simeq-0.5$ is most appropriate for  the UDF sample where $\sigma_t\equiv\sigma(2\Re)$~\footnote{When using the alternative definition of $\sigma_t\equiv\sigma_0$ (discussed in \S~\ref{section:methodo:3D}) used in \citet{UblerH_19a}, we find that  $q\simeq -0.7$ to $-0.8$ very similar to the value of $-0.8$ we derived from the \citet{UblerH_19a} data (their Fig.~6).}.



 
 \begin{table*}
     \centering
     \begin{tabular}{cccccccccc}
     Data & Model & 
 $\alpha$ &
$\beta$ &
$\gamma$ &
$a$ &
$b$ &
$q$ \\
\hline
(1) & (2) & (3) & (4) & (5) & (6) & (7) & (8)   \\
\hline
  UDF & Fiducial &  
 $\FITalpha$ &
 $\FITbeta$ &
 $\FITgamma$ &
 $\FITzalpha$ &
 $\FITZP$  & -0.5  \\ 
   TNG50 & Fiducial & 
 -0.03 $^{+0.05}_{-0.05}$ & 
 0.18 $^{+0.07}_{-0.07}$ & 
 1.28 $^{+0.10}_{-0.10}$ & 
 -0.45 $^{+0.24}_{-0.27}$ &  
 2.54 $^{+0.03}_{-0.04}$ & 
 -0.87
\\
TNG50 & noVS & 
   0.05 $^{+0.09}_{-0.09}$ &
  0.53 $^{+0.08}_{-0.12}$ &  
--- & 
  +0.06 $^{+0.28}_{-0.28}$ &  
  2.78 $^{+0.03}_{-0.03}$ & --- \\
  \hline
     \end{tabular}
     \caption{Fitted parameters to $\js-M_\star$ from Eq.~\ref{eq:Jtot:fit}.
     (1) Data set;
     (2) Model name;
     (3-7) Parameters from Eq.~\ref{eq:Jtot:fit} with 95\%\ confidence levels (2$\sigma$);
     (8) Redshift evolution of $V/\sigma_{\rm t}$ (Eq.~\ref{eq:VoverS:evolution}).
      }
      \label{tab:Jtot:fit}
 \end{table*}
 
 \subsubsection{The $j-M-(V/\sigma)$ global fit}

 We use  a Bayesian algorithm to fit the 5 parameters ($\alpha,\beta,\gamma,a,b$) using   a Student-$t$ likelihood~\footnote{ A Student-$t$ likelihood  is preferred over a Gaussian likelihood because it is naturally robust against potential outliers from its larger tail.} with
 the NoU-Turn Sampler (NUTS) from the \textsc{PyMC3} python package \citep{pymc3}.
 In Fig.~\ref{fig:Jtot}, we show the resulting  fit to the $j-M-(V/\sigma_{\rm t})$ relation  using Eq.~\ref{eq:Jtot:fit} whose fitted parameters are listed in  Table~\ref{tab:Jtot:fit}.
From this fit, one sees that the $\js-M_\star$ relation is a nonlinear function of $M_\star$, as the $\js-M_\star$ relation flattens toward low masses given that the quadratic factor $\alpha$ is significantly non zero ($\alpha=0.04\pm0.01$). Specifically, the slope of the sAM relation $s\equiv{\rm d}\log \js/{\rm d}\log M_\star$ goes from $ s\approx0.5$ at $\log M_\star/$M$_\odot=10.5$, consistent with numerous results at $z=0$ \citep[e.g.,][]{ObreschkowD_14a,CorteseL_16a,PostiL_18b,FallM_18a}  or high-redshifts \citep[e.g.,][]{SwinbankM_17a,MarascoA_19a},  to $s\approx0.25$ in the low mass regime at $\log M_\star=8$. In the remainder of this paper, 
we   use $f(\log M)$ as short hand for this nonlinear mass dependence in Eq.~\ref{eq:Jtot:fit} for the remainder of this work.

 This nonlinear $j-M$ relation is expected from  Eq.~\ref{eq:js:formula} where the stellar mass fraction $f_m$ plays a significant role in this mass regime.
Together, this result indicates the angular momentum retention fraction $f_j$  at  $z\sim1$ is not far from being constant with mass   \citep[see][for discussions on this at $z=0$]{PostiL_18a,PostiL_19a}.

 \subsubsection{The $j-M$ redshift evolution}
 \label{section:UDF:sAM:redshift}
 
\begin{figure*}
 \centering
 \includegraphics[width=\textwidth,height=8cm]{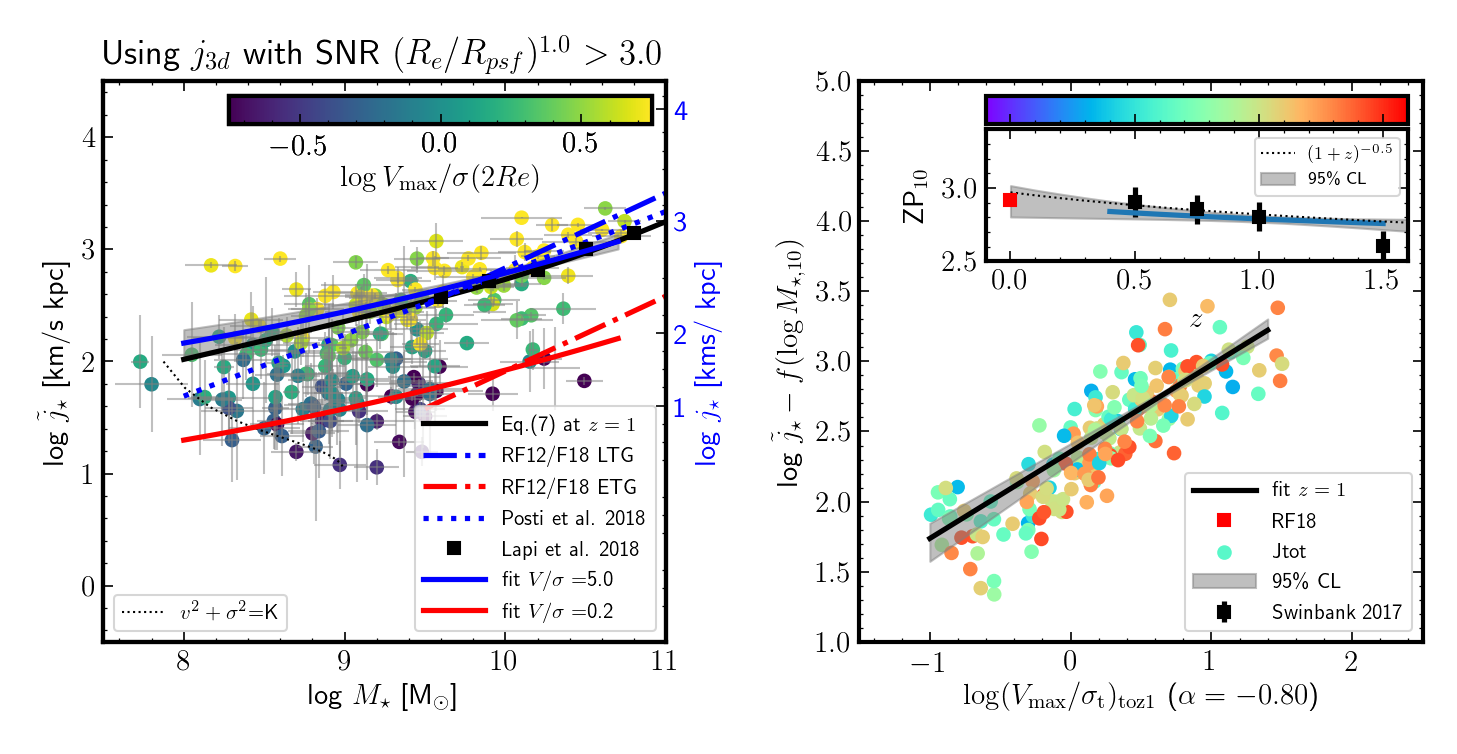}
 \caption{Specific angular momentum for the \Nsamp\ galaxies in the MUSE UDF mosaic fields with a max S/N $\gtrsim3$.
 {\bf Left: } Total specific angular momentum $\js$ (SFR weighted) as a function of stellar mass $M_\star$.
 The points are color-coded as a function of $V/\sigma_{\rm t}$.
 The right-axis shows the corresponding $j_\star$ according to the 0.25dex offset found in Fig.~\ref{fig:Jcomparison}.
 The blue (red) dashed line show the $j_\star\propto M_\star^{0.6}(0.75)$ sequence for late (early) type galaxies at $z=0$ from \citet[][FR18]{FallM_18a}.  
 The dotted line (solid squares) show the $z=0$ sAM $j_\star-M_\star$ sequence from 
 \citet{PostiL_18b} \citep{LapiA_18a}, respectively.
 The dotted line show the $j_\star\propto M_\star^{0.55}$ sequence for $z=0$ disk galaxies of \citet{PostiL_18b}.
 The solid black line shows the expected $j_\star-M_\star$ relation from Eq.~\ref{eq:js:predict}.
The solid blue and red lines show the fitted fiducial model from Eq.~\ref{eq:Jtot:fit} and the gray band represents the 95\%\ confidence interval.
 {\bf Right: } The normalized AM sequence $\js/f(\log M_\star)$ as a function of   $(V/\sigma_{\rm t})_{\rm toz1}$ (normalized to $z=1$ see text), i.e. after taking into account for the redshift evolution of $\sigma_{\rm t}(z)$,  where the points are color-coded with redshift.
In the inset, we show the redshift evolution of the sAM relation at $M_\star=10^{10}\Msun$
 (solid line)  at fixed $(V/\sigma_{\rm t})_{\rm toz1}$ (=5) along with  the 95\%\ predictive interval (gray band), 
with the $z=0$ ZP from \citet{FallM_18a} (red square), the  \citet{SwinbankM_17a} binned data (see text), and the
expected $(1+z)^{-0.5}$ evolution (dotted line) from Eq.~\ref{eq:js:predict}.
}
 \label{fig:Jtot}
 \end{figure*}

 In regards to the redshift dependence of the sAM sequence, the inset in Fig.~\ref{fig:Jtot}(b) 
 shows the redshift evolution of the zero-point ZP$_{10}$ (Eq.~\ref{eq:Jtot:fit})  of the $\js-M_\star$ relation  fitted over our redshift range $0.4<z<1.4$ for $M_\star=10^{10}\Msun$ and $V/\sigma_{\rm t}=5$.
  The red circle shows the RF12 $z=0$ zero-point, and the squares show the \citet{SwinbankM_17a} results~\footnote{For \citet{SwinbankM_17a}, we assumed a mean $V/\sigma_{\rm t}$=2.5 for their $V/\sigma_{\rm t}>1$ sample according to their sample properties. 
  For \citet{RomanowskyA_12a}, we assumed a $V/\sigma=5$ for their disk-dominated sample and shifted their  $j_\star$ zero-point to our SFR-weighted $\js$ by -0.25~dex.}. 
  The dashed line shows the $(1+z)^{-1/2}$ expected evolution (Eq.~\ref{eq:js:formula}). This panel shows that the sAM relation (at fixed $V/\sigma_{\rm t}$)
  is evolving   as $$\js/f(\log M_\star)\propto(1+z)^{a},$$ with $a=\FITzalpha$ (95\%) which is  consistent with the theoretical evolution $(1+z)^{-1/2}$ from Eq.~\ref{eq:js:formula}.
 This redshift dependence is not consistent  with the slope of $-1$ found by  
\citet{SwinbankM_17a} to $z=1.5$, but it should be noted that they did not perform a global fit to their data and that their binned data (squares in the inset of Fig.~\ref{fig:Jtot}b) are in fact  consistent with a slope of $-0.5$ up to $z=1.0$, with the $z=1.5$ binned data not matching the $(1+z)^{-0.5}$ trend
is  most likely due to incompleteness issues at these redshifts.

While \citet{SwinbankM_17a} and \citet{HarrisonC_17a}  found some evidence for a redshift evolution of the $j-M$ relation with a combined data-set of $\sim$1000 SFGs, in contrast,
\citet{MarascoA_19a} found no evidence for a redshift evolution of the $\js-M_\star$ relation on a subset (made of 17 $z=1$ disk galaxies with $M_\star>10^{9.5}~\Msun$) of the KROSS and KMOS3D survey. This discrepancy implies no evolution of both the TFR relation and of the size-mass relation, in apparent contradiction to several studies \citep[e.g.,][]{StraatmanC_17a,UblerH_17a,vanderWelA_14a}. This discrepancy could originate from the small sample size, from the different methodologies involved, and/or could be resolved under our $(V/\sigma_{\rm t}$) framework.

 \begin{figure*}
 \centering
 \includegraphics[width=\textwidth,height=8cm]{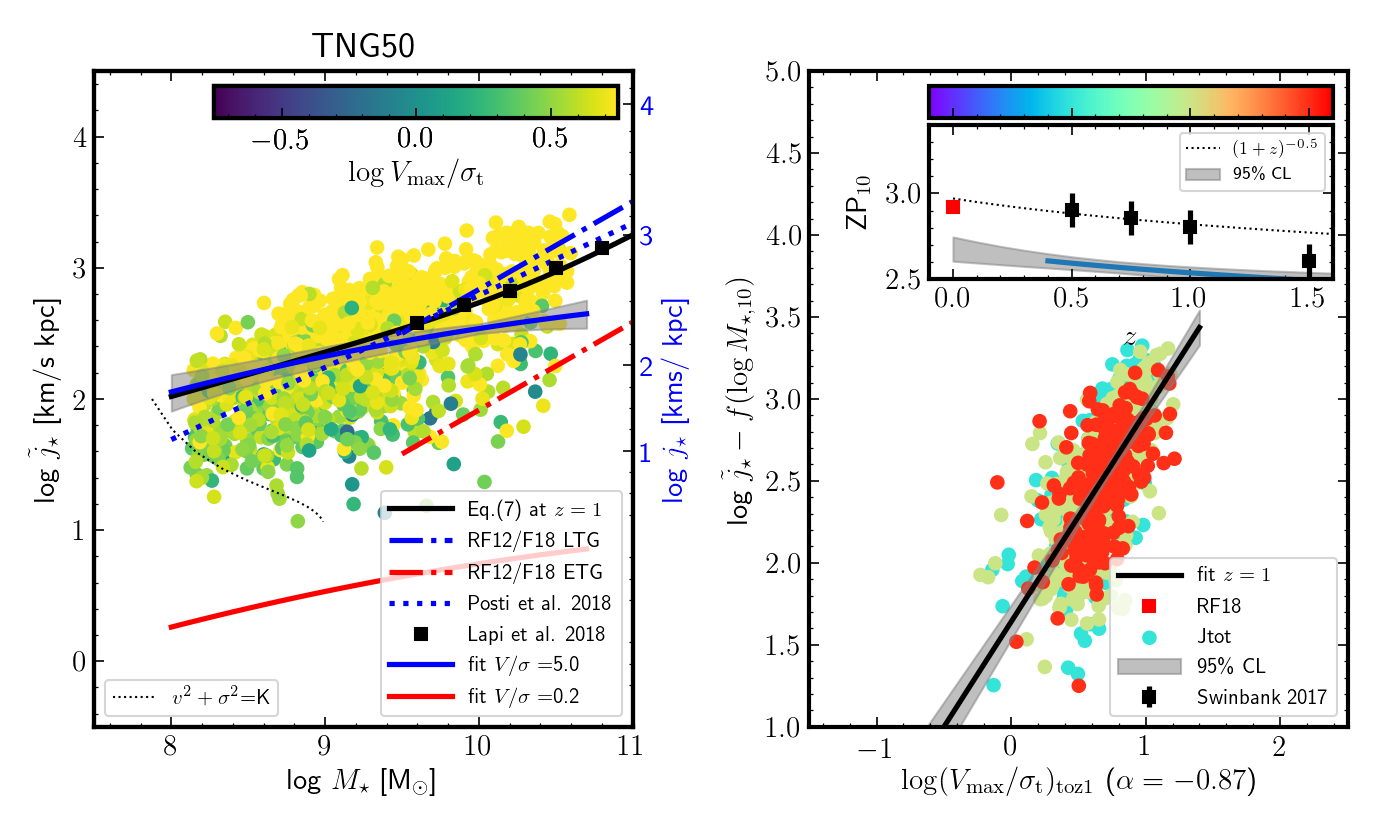}
 \caption{Same as \Fig{fig:Jtot}, but for TNG50 SFGs. }
 \label{fig:Jtot:TNG}
 \end{figure*}

\section{Comparisons to SFGs in TNG50}
\label{section:tng50:comparison}

In order to compare our results on the sAM relation presented in \S~\ref{section:udf:sAM} to TNG50, we selected 18000 SFGs at $z=0.5$, 1.0 and 1.5 from TNG50. We matched the properties of TNG50 SFGs to the UDF SFGs as follows. We first match the SFR distribution by imposing a $\log$(SFR/[M$_\odot$~yr$^{-1}])>-1.5$ as well as impose a similar observational limit of $\sqrt{V_{\rm max}^2+\sigma_{\rm t}^2}>30\kms$, and then match both the redshift and stellar mass distributions of the UDF sample. The resulting sample is made of 5180 SFGs.

Fig.~\ref{fig:Jtot:TNG}(a)
shows the SFR-weighted sAM $\js$ as a function of $M_\star$ for  SFGs in TNG50 with $M_\star>10^8\Msun$, color-coded with $(V/\sigma_{\rm t})$.
Fig.~\ref{fig:Jtot:TNG}(b) shows the relation between the normalized sAM $\js/f(M_\star)$ and $(V/\sigma_{\rm t})_{\rm toz1}$, that is  $V/\sigma_{\rm t}$ corrected for its redshift evolution to $z=1$ using the scaling $V/\sigma_{\rm t}\propto(1+z)^{0.87}$ found in \Fig{fig:appendix:VoverSigma}.
In this figure, we show only 1000 SFGs for clarity.
Fig.~\ref{fig:Jtot:TNG} shows that SFGs in TNG50 follow  similar trends with $M_\star$ and with $V/\sigma_{\rm t}$ as in the UDF sample.
In particular,  the $\js-M_\star$ relation is also a strong function of $V/\sigma_{\rm t}$.

To quantify potential differences, we applied the same parametric fit as for the UDF sample on a randomly selected subset of 250 galaxies in order to have  statistical errors similar to the UDF sample, namely using  \Eq{eq:Jtot:fit} with the redshift evolution of $V/\sigma_{\rm t}$ (\Eq{eq:VoverS:evolution}).
 The results are listed in Table~\ref{tab:Jtot:fit} with and without the  correction for the redshift evolution of $V/\sigma_{\rm t}$.
   One sees that the $V/\sigma_{\rm t}$ dependence is  stronger in TNG50 than in the UDF sample with $\gamma\sim1.3$, and the redshift evolution for the TNG50 SFGs is consistent with the $(1+z)^{-0.5}$ expectation,  with $a=-0.45\pm0.25$, against $a\simeq \FITzalpha$ for the UDF SFGs.
 
We end this section by noting that we found no strong dependence of the sAM with age (formation redshift) or with morphological indicators \citep{TacchellaS_19a}, such as Gini-M20, and bulge-to-total ratio $B/T$.


\begin{figure*}
\centering
\includegraphics[width=\textwidth]{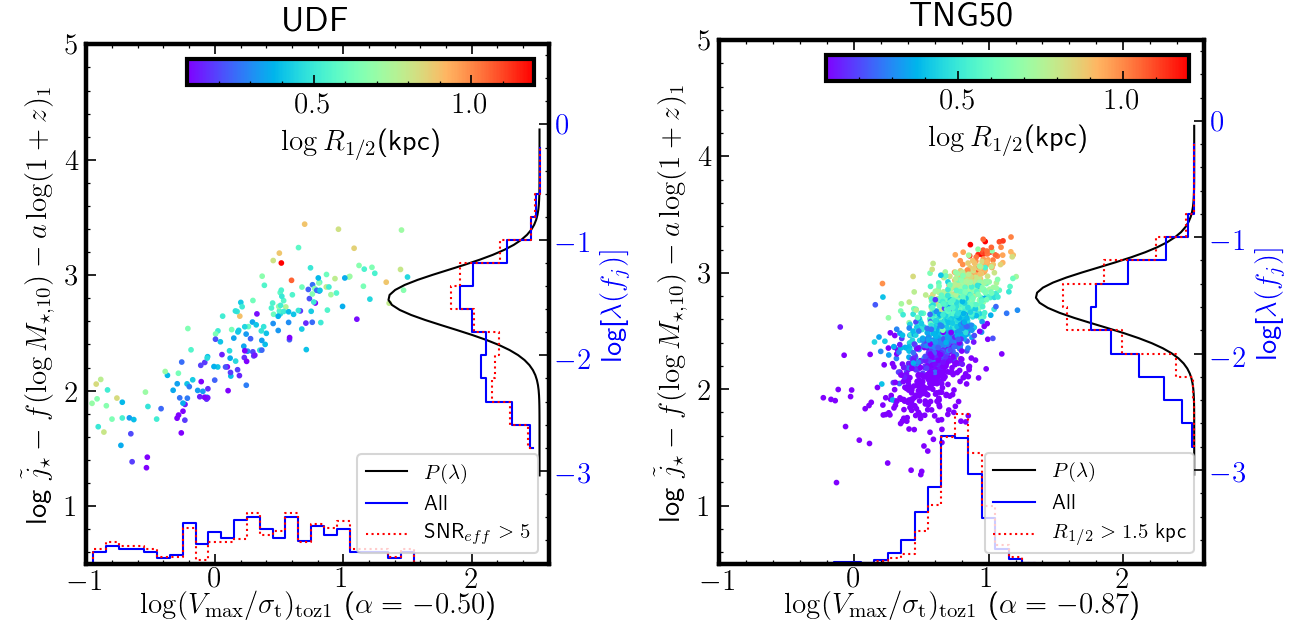}
\caption{{\bf Left:} The sAM sequence normalization as a function of ($V/\sigma_{\rm t})_{\rm toz1}$ for the UDF mosaic sample of SFGs where the color representes the SFR-weighted $\log R_{1/2,\OII}$.
The left $y$-axis $\log \js-f(\log M_\star)-g(z)$ is found from the fiducial fits shown in Figs.~\ref{fig:Jtot}-\ref{fig:Jtot:TNG} (see Table~\ref{tab:Jtot:fit}).
The right $y$-axis shows the corresponding effective spin $\lambda \times f_j$ according to Eq.~\ref{eq:js:predict} using the zero-poing of Eq.\ref{eq:js:predict}.
The solid histograms show the distributions from the data. The solid line shows the log-normal distribution  $P(\lambda)$ expected for halos \citep{BettP_07a}.
{\bf Right:} Same for the matched TNG50 sample of SFGs. 
}
\label{fig:Jtot:spin}
\end{figure*}

 \section{The $j-M-(V/\sigma)$ relation: implications}
 \label{section:implications}

 As shown in Fig.~\ref{fig:Jtot} and Fig.~\ref{fig:Jtot:TNG}, the sAM sequence of SFGs in the UDF and in TNG50 represents a strong function of the galaxy dynamical state as characterized by  $V/\sigma_{\rm t}$, where rotation dominated SFGs with $V/\sigma_{\rm t}\sim5$ and dispersion dominated SFGs with $V/\sigma_{\rm t}=0.1$ are offset by $\sim1$dex. This result re-inforces the preliminary results of \citet{ContiniT_16a}
  on only 27 intermediate redshift SFGs, and extends the similar trend present in \citet{BurkertA_16a} for massive (with $M_\star>10^{10}\Msun$) SFGs.

This $j-M-(V/\sigma_{\rm t})$ relation has important consequences.
If SFGs form  with high sAM, on the rotation-dominated $j-M$ sequence, this $j-M-(V/\sigma_{\rm t})$  relation indicates that SFGs may experience a dynamical  transformation to lower their sAMs  prior to becoming passive, and remain on the early-type $j-M$ sequence.
This might also lead to  a morphological transformation as the
$j-M$ residuals also correlate with \sersic{} $n$ \citep{HarrisonC_17a}, but this trend is not found in the UDF data. Indeed, the vast majority of our SFGs have \sersic{} indices $n\lesssim1$, thus they have not undergone the morphological transformation related to bulge formation. 
This indicates that the kinematical transformation (from high $V/\sigma_{\rm t}$ to low $V/\sigma_{\rm t}$) occurs before the morphological transformation toward bulge formation. 

This scenario of  kinematical transformation toward low $j$ with smaller sizes~\footnote{In this context, we note that our SFGs with low $V/\sigma_{\rm t}$ also tend to be smaller (as discussed below). } and larger dispersions  is in agreement with the recent finding  from the CANDELS survey \citep{OsborneC_20a} that  SFGs contains $\sim15$\%{} of compact objects ('blue nuggets'),
 indicating that compaction precedes quenching. 


In local galaxies, \citet{OhS_20a} separated the bulge and disk (stellar) kinematics of galaxies  in the SAMI sample and found that the bulge  (disk) component have low (high) $V/\sigma$, and follows the Faber-Jackson (Tully-Fisher) relation, respectively. We hypothesize that our $0.4<z<1.4$ SFGs are their progenitors, and $z=1$ SFGs with low $V/\sigma$ will evolve into bulge dominated systems at $z=0$.  
This is consistent with the TNG results of \citet{TacchellaS_19a}, who found that the morphology of galaxies is  set during their star-forming phase.

We return to Eq.\ref{eq:js:predict} to gain insight on our results and in particular to address whether our relatively broad sAM sequence is consistent with the expected distribution of halo spins $\lambda$. One might expect the distribution of halo spins to be reflected in the scatter of the sAM sequence if each SFG retains a similar amount of sAM (a constant $f_j=1$), as in semi-analytical models. However, in hydro-dynamical simulations the situation is more complex \citep{GenelS_15a,JiangF_19a} given the potential large AM exchanges.

\Fig{fig:Jtot:spin} shows the normalization of Eq.\ref{eq:js:predict}, namely $\log \js-f(\log M_\star)-a\log(1+z)$, as a function of $V/\sigma$ (corrected again to $z=1$) for the MUSE sample (left) and for the TNG50 sample (right), respectively. The left $y$-axis  is found from the ``fiducial'' fits shown in Figs.~\ref{fig:Jtot} (\ref{fig:Jtot:TNG}), which are listed in Table~\ref{tab:Jtot:fit}. 
The histograms in \Fig{fig:Jtot:spin} show  the distributions of Eq.\ref{eq:js:predict} recast as $\lambda f_j$ (right $y$-axis). For comparison, the solid line shows the log-normal distribution  $P(\lambda)$ expected for dark-matter halos \citep{BettP_07a} assuming a retention factor $f_j$ of unity.

This figure shows that the distribution of SFGs in sAM is consistent with the expectation from dark-matter halos in TNG50, but is broader in our observations. The trend with $V/\sigma$ is a strong function of galaxy sizes (SFR-weighted) in TNG50, but not in the MUSE sample.
{Another difference is the paucity of dispersion dominated galaxies in TNG50 (with 15\%\ having $V/\sigma<3$) compared to the MUSE sample (with 50\%\ having $V/\sigma<3$).}
Interestingly, the MUSE sample restricted to S/N$_{\rm eff}>5$ appears to show a more bimodal distributions in $\lambda f_j$, consistent with the $z=0$ RF12 results where rotation dominated galaxies have retained most of their halo sAM, while dispersion dominated systems have lost around 80\%\ of their sAM.

\section{Conclusions}
\label{section:conclusions}

The main purpose of  this paper was to develop a robust methodology to measure $j_{\rm tot}$ from the angular momentum profiles $j(<r)$ of galaxies in the low-mass regime (down to $M_\star=10^{8}\Msun$), even when one is severely limited by S/N, surface-brightness limits and/or by the spatial resolution of ground-based instruments.
The methodology consists of a two-step process (\S~\ref{section:methodo:3D}) and is based on the fitting algorithm \galpak{} \citep{BoucheN_15a}. The algorithm adjusts the morphological and dynamical parameters of a 3-dimensional disk model against the IFU data directly. 

 Using mock observations of SFGs taken from the TNG50 simulations \citep{NelsonD_19a,PillepichA_19a} with $M_\star=10^{8.5-10.5}\Msun$, we have shown that
 our 3D methodology  is able to recover the angular momentum profiles of SFGs with little bias ($<5$\%)  (Fig.~\ref{fig:dJ}) provided the S/N and/or surface brightness is sufficient (see~\S~\ref{section:tng50:results}). 
 In particular, we find that we are able to estimate the $j$ profiles
 of low-mass SFGs down to $10^8\Msun$ with correspondingly small sizes. Specifically,
 galaxies must have  S/N$_{\rm eff}\gtrsim3$, where S/N$_{\rm eff}=$S/N$_{\rm max}\,(\Re/R_{\rm psf})$ and S/N$_{\rm max}$ is the S/N in the brightest pixel, $\Re$ is the half-light radius and $R_{\rm psf}$ is the PSF radius.

Using our 3D methodology on a sample of \Nsamp\ SFGs (selected solely from their \OII{} emission) with $M_\star>10^8\Msun$ at $0.4<z<1.5$ from the 9~arcmin$^2$ MUSE observations of \citet{BaconR_17a}, we have found  the following: \\

$\bullet$ Our methodology allows us to determine galaxy sizes of unresolved SFGs with down to $R_{1/2,\OII}\simeq1$ kpc or 0.2" (Fig.~\ref{fig:Re})
thanks to the built-in use of the PSF in our modeling. This opens a new parameter space (Fig.\ref{fig:size:redshift}b) between $1<\Re<3$ kpc (and $10^8<M_\star<10^{9.5}\Msun$) unexplored so far with IFS surveys  beyond the local universe
\citep[e.g.,][]{BurkertA_16a,SwinbankM_17a,HarrisonC_17a}.

$\bullet$ We find that the $\js-M_\star$  relation for SFGs is a nonlinear relation (\Fig{fig:Jtot}a) with $\js\propto M_\star^{0.3-0.4}$ at $M_\star<10^{9.5}\Msun$ and $\js\propto  M_\star^{0.6}$ at $M_\star>10^{9.5}\Msun$, similarly to RF12.
This quadratic $j-M$ relation appears to follow the expectations from simple disk formation models with the observed stellar mass fractions $f_m$ \citep{MosterB_10a,BehrooziP_13b}.
The zero-point of the sAM sequence at $z=1$ is found to be $\log \js = 2.85\pm0.05$  for rotation dominated  $\log M_\star=10$ SFGs both in the UDF and TNG50 samples.

$\bullet$ The $\js-M_\star$ relation is a strong function of the dynamical status of SFGs quantified by $V/\sigma_{\rm t}$, leading to a $\js-M_\star-(V/\sigma)$ sequence (\Fig{fig:Jtot}b), where rotation dominated SFGs (with $V/\sigma_{\rm t}>1$--5) follow the $j-M$ relation of local disks, while dispersion dominated galaxies (with $V/\sigma_{\rm t}=0.1$) follow the $j-M$ relation of early types.  This confirms the preliminary results of \citet{ContiniT_16a} on a small sample and extends the results of \citet{BurkertA_16a} to the relatively unexplored regime below $10^{10}\Msun$ down to $10^ 8\Msun$.
 
 $\bullet$ This $j-M-(V/\sigma_{\rm t})$ relation is reminiscent of the $z=0$
 results where the $j-M$ relation is found to be a function of the bulge-to-total  $B/T$ ratio in local galaxies \citep[e.g.,][]{ObreschkowD_14a,FallM_18a}
and is in relatively good qualitative agreement with  theoretical framework on AM loss increasing with B/T in the EAGLE simulations \citep{ZavalaJ_16a,LagosC_17a} and with the results of \citet{GenelS_15a} from the Illustris simulations. 
Consequently,  SFGs   experience a dynamical transformation (from the disk to early type sequence of the $j-M$ relation) and lower their  sAMs  prior  to  becoming  passive,  on  the early-type $j-M$ sequence.

$\bullet$ The ZP of the $j-M-(V/\sigma_{\rm t})$ relation in the UDF sample (\Fig{fig:Jtot}b) evolves as $\propto(1+z)^{a}$ with $a=\FITzalpha$ (2$\sigma$), which is consistent with the expectation of $(1+z)^{-0.5}$, when taking into account the redshift evolution of $V/\sigma_{\rm t}$ ($\propto(1+z)^{-0.5}$) for our definition of $\sigma_{\rm t}=\sigma(2\Re)$.

$\bullet$  The ZP of the $j-M-(V/\sigma_{\rm t})$ relation in the TNG50 sample (\Fig{fig:Jtot:TNG}b) evolves as $\propto(1+z)^{-0.45\pm0.25}$, which is consistent with the expectation of $(1+z)^{-0.5}$, after taking into account the redshift evolution of $V/\sigma_{\rm t}$, which goes as $\propto(1+z)^{-0.85}$ for our definition of $\sigma_{\rm t}=\sigma(2\Re)$. The  $j-M$ relation in the TNG50 sample is steeper with $V/\sigma_{\rm t}$. This is likely to a combination of the difficulty in measuring $\sigma_t$ in observations down to 10$\kms$ and to the physics detail in TNG50.

 $\bullet$ The galaxy $\lambda$ spin parameters (from the scatter of the $j-M-(V/\sigma)$ relation) is broader in the MUSE sample than in TNG50 (\Fig{fig:Jtot:spin}), and  also correlates better with galaxy sizes (SFR-weighted) in TNG50 than in the UDF observations.

$\bullet$ The MUSE sample restricted to S/N$_{\rm eff}>5$ appears (\Fig{fig:Jtot:spin}a) to show a more bimodal distributions in sAM, consistent with the $z=0$ RF12 results where rotation dominated galaxies have retained most of their halo sAM, while dispersion dominated systems have lost around 80\%\ of their sAM.

We have shown that both in the UDF MUSE mosaic and in TNG50, SFGs follow a $j-M-(V/\sigma)$ fundamental relation, and have highlighed the importance of taking into account  the redshift evolution of $V/\sigma$  in the redshift evolution of the sAM sequence $j-M$.

Our results demonstrate that it is possible to study the kinematics of large samples of SFGs  blindly (i.e., without preselection on morphology, nor prior {\it HST} continuum detection) with little biases including the low-S/N regime of 3 to 5. Our 3D modeling technique opens the low-mass regime, $M_\star<10^9\Msun$, at intermediate redshifts $0.5<z<1.5$, which had been 
unexplored so far.


\section*{Acknowledgments}
This  work  made  use  of  the  following  open  source
software:  matplotlib \citep{matplotlib}, 
NumPy \citep{numpy}, 
SciPy \citep{scipy}, 
Astropy  \citep{astropy2018}.

We thank the referee for his/her constructive comments that improved  the  quality  of  the  paper.
We thank L. Posti for insightful discussions that led to an improved manuscript.
This work has been carried out thanks to the support of  the ANR 3DGasFlows (ANR-17-CE31-0017), the OCEVU Labex (ANR-11-LABX-0060). 
B.E. acknowledges financial support from
the program National Cosmology et Galaxies (PNCG) of CNRS/INSU with INP and IN2P3, cofunded by CEA and CNES. 
R.B. acknowledges support
from the ERC advanced grant 339659-MUSICOS.
S.G., through the Flatiron Institute, is supported by the Simons Foundation.
The primary TNG simulations were realized with compute time granted by the Gauss Centre for Super- computing (GCS): TNG50 under GCS Large-Scale Project GCS- DWAR (2016; PIs Nelson/Pillepich), and TNG100 and TNG300 under GCS-ILLU (2014; PI Springel) on the GCS share of the supercomputer Hazel Hen at the High Performance Computing Center Stuttgart (HLRS). GCS is the alliance of the three national supercomputing centres HLRS (Universit\"at Stuttgart), JSC (Forschungszentrum J\"ulich), and LRZ (Bayerische Akademie der Wissenschaften), funded by the German Federal Ministry of Education and Research (BMBF) and the German State Ministries for Research of Baden-W\"urttemberg (MWK), Bayern (StMWFK) and Nordrhein-Westfalen (MIWF).

\appendix

\section{Analytical derivation }
\label{sec:appendix:analytic}

 For a \sersic\ profile $I(r)\propto \exp\left[{-b_n\,(R/R_{\rm e})^{1/n}}\right]$, we can calculate the coefficient $k_n$ in \Eq{eq:RF12}  assuming a constant flat rotation curve such as $V(R)=V_{\rm c}$ as in RF12.  
 Using Eq.~\ref{eq:AM:2d}  and by substituting by $x=b_n\left(\frac{R}{Re}\right)^{1/n}$, we find
\begin{align}
j_{\rm t}&=V_{\rm c}\frac{\int{\rm d}R~R^2~{\rm e}^{-b_n\left(\frac{R}{R_{\rm e}}\right)^{1/n}}}{\int{\rm d}R~R~{\rm e}^{-b_n\left(\frac{R}{R_{\rm e}}\right)^{1/n}}},\\
&=V_{\rm c}\frac{n\frac{R_{\rm e}^3}{b_n^{3n}}\Gamma(3n)}{n\frac{R_{\rm e}^2}{b_n^{2n}}\Gamma(2n)},\\
&=V_{\rm c}R_{\rm e}\frac{1}{b_n^n}\frac{\Gamma(3n)}{\Gamma(2n)},
\end{align}
which means that the $k_n$ coefficient in \Eq{eq:RF12} is
\begin{equation}
\label{eq:kn}
k_n = \frac{1}{b_n^n}\frac{\Gamma(3n)}{\Gamma(2n)}.
\end{equation}
where $\Gamma$ is the gamma function.
As discussed in \citet{GrahamA_05a}, the \sersic\ coefficient $b_n$ is given from the solution of the following equation
\begin{equation}
\Gamma(2\,n)=2\,\gamma(2\,n,b_n)\label{eq:bn:exact}
\end{equation}
where $\Gamma$ is the (complete) gamma function and $\gamma$ is the incomplete gamma function. Note, 
the \sersic\ coefficient $b_n$ can be also approximated by  $b_n\approx1.9992\,n-0.327 \label{eq:bn:approx}$
as discussed in \citet{GrahamA_05a} \cite[see also][]{PrugnielP_97a}.

 \section{Selection function}
 \label{sec:appendix:limit}
 
 Due to the instrumental resolution of $R\sim3000$  ($\sim100\kms$ or $\sim50 \kms$ per pixel) for MUSE, we are limited to a few tens of $\kms$ in either $V_{\rm max}$ or $\sigma_{\rm t}$. In order to quantify the impact of this limit to the 
   $j-M-(V/\sigma)$ relation, $\log\, \js=f(\log M_\star)+\gamma\log(V_{\rm max}/\sigma_{\rm t})$, we assume a constant floor $K$ for $V_{\rm max}^2+\sigma_{\rm t}^2$, which becomes
    \begin{eqnarray}
 K&=r^2\left(\sin^2(\theta)+\cos^2(\theta)\right)
 \end{eqnarray}
 after a change of variables with $\tan(\theta)=\frac{v}{\sigma_{\rm t}}$ and $|K|=r^2$.
This implies that  the $j-M-(V/\sigma)$ relation becomes
 \begin{eqnarray}
     \log\, \js&=&f(\log M_\star)+\gamma\log(\tan(\theta)).
 \end{eqnarray}
Finally, the last term can be expressed as a function of $M_\star$ using the S05-$M_\star$ relation \citep[e.g.,][]{KassinS_07a,UblerH_17a,TileyA_19a}, namely $\log$~S05$\simeq 0.3 \log M_{\star,9} + 1.5$, and
S$^2_{05}\equiv 0.5 V_{\rm max}^2+\sigma_{\rm t}^2=r^2\,\left[0.5\sin^2(\theta)+\cos^2(\theta)\right].$
  
 \section{Kinematics}

\begin{figure*}
\includegraphics[width=\textwidth]{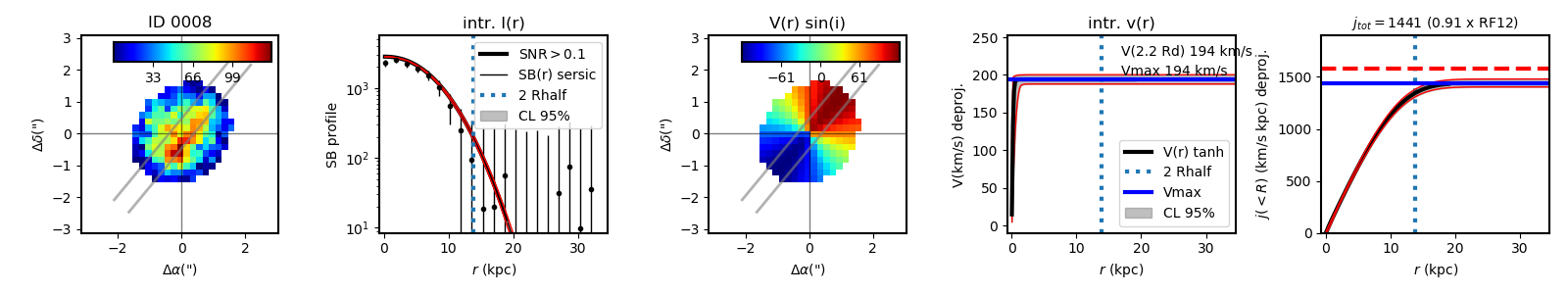}
\includegraphics[width=\textwidth]{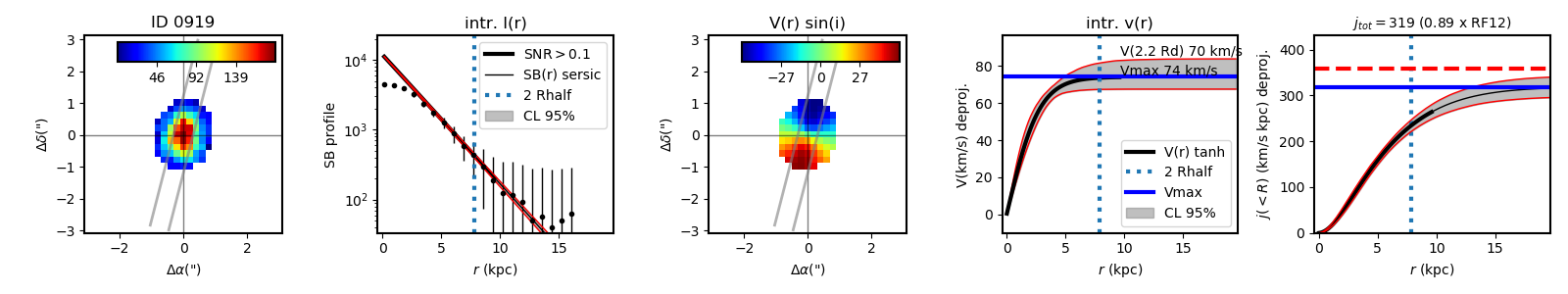}
\includegraphics[width=\textwidth]{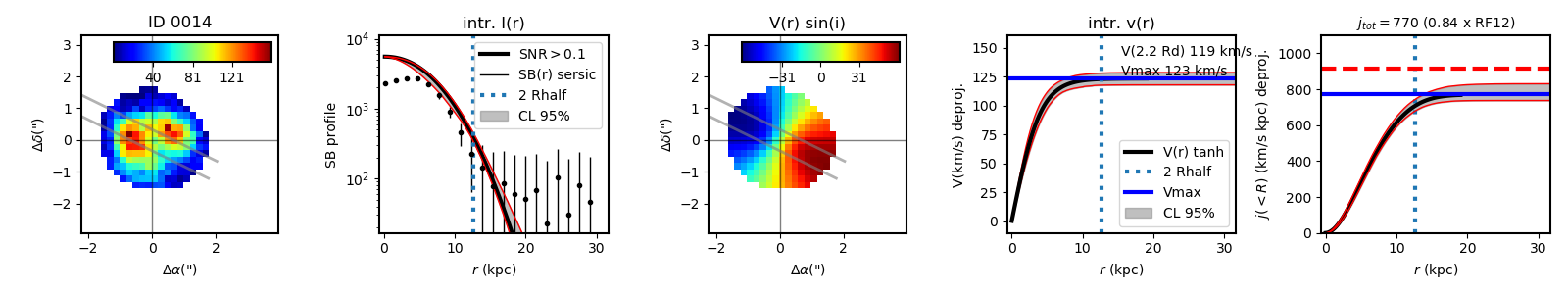}
\includegraphics[width=\textwidth]{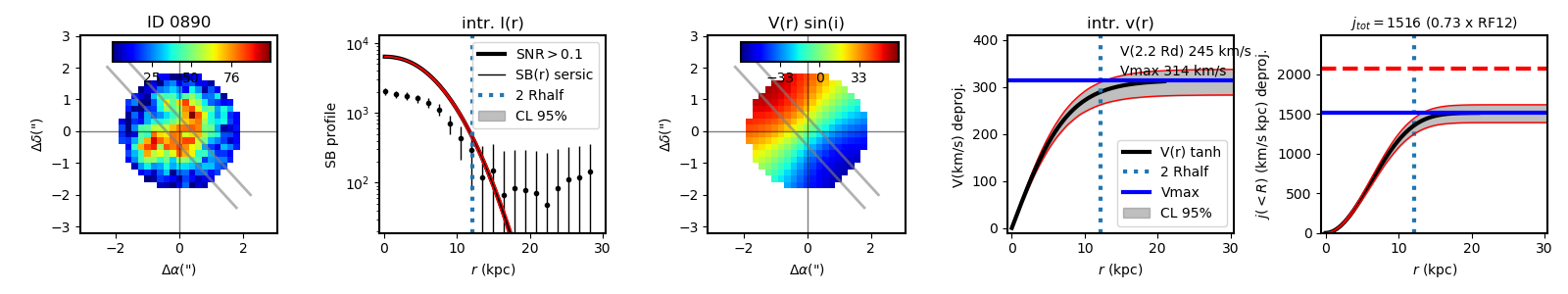}
\includegraphics[width=\textwidth]{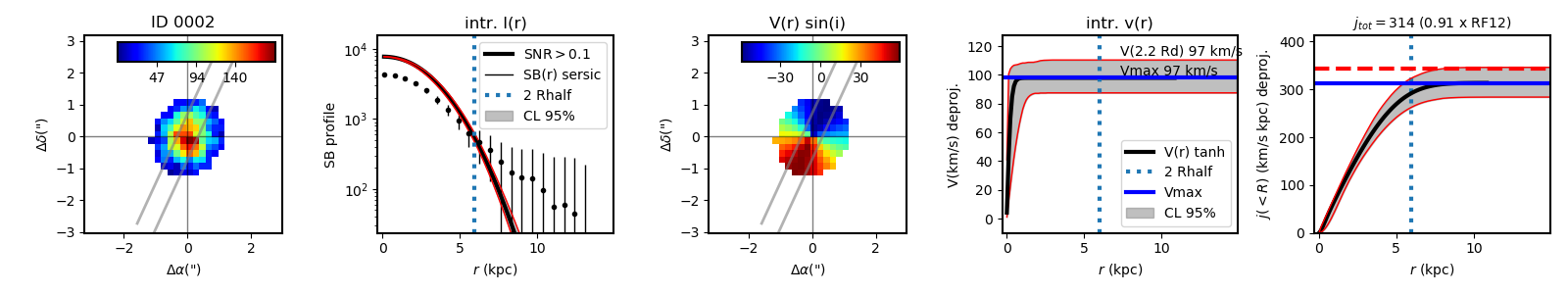}
\caption{Morphology and kinematics of examples from the UDF-mosaic sample.
The panels show the flux map, the SB profile $\Sigma_{\rm SFR}$, the velocity field, the rotation curve and the SFR-weighted $\js(<r)$ profiles, respectively.
The 1D profiles are determined along the pseudo-long slits represented by the gray straight lines (black solid lines), along with the 95\%\ confidence intervals. The vertical dotted lines represent $2\times\Rhalf$.
In the last column, the horizontal blue lines show the total modeled angular momentum $j_{\rm 3D}$, while the horizontal red lines show the total $\js$ using the  \Eq{eq:RF12} approximation.
\label{fig:udfmosaic:examples}}
\end{figure*}

 Fig.~\ref{fig:udfmosaic:examples} shows examples of kinematic properties for UDF mosaic galaxies analyzed in \S~\ref{section:udf:sAM}.
 
\begin{figure*}
\centering
\includegraphics[width=\textwidth]{figs/sub315171_AMprofile.png}
\includegraphics[width=\textwidth]{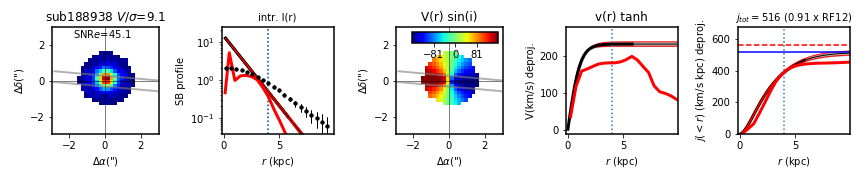}
\includegraphics[width=\textwidth]{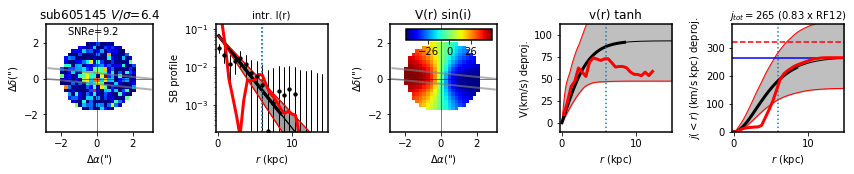}
\includegraphics[width=\textwidth]{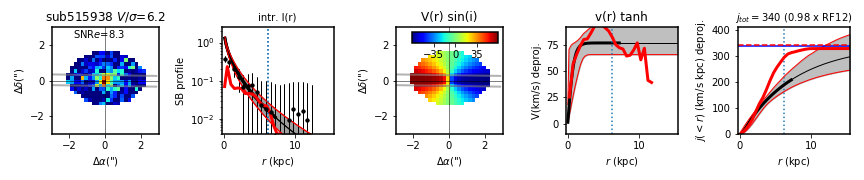}
\includegraphics[width=\textwidth]{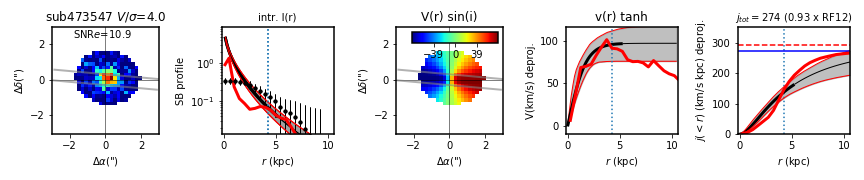}
\includegraphics[width=\textwidth]{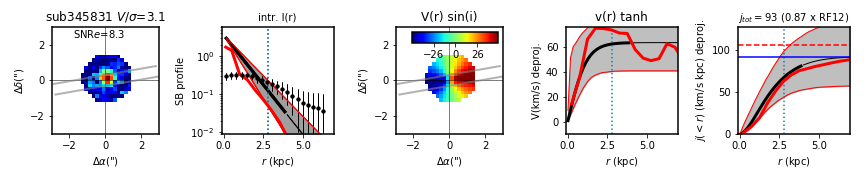}

\caption{Kinematic properties of some of the TNG50 galaxies. The panels show the flux map, the SB profile $\Sigma_{\rm SFR}$, the velocity field, the rotation curve and the SFR-weighted $\js(<r)$ profiles, respectively.
The 1D profiles are determined along the pseudo-long slits represented by the gray straight lines (black solid lines), along with the 95\%\ confidence intervals. The vertical dotted lines represent $2\times\Rhalf$.
The 1D red solid lines represent the true profile determined directly from the TNG50 data. 
 In the last column, the horizontal blue lines show the total modeled angular momentum $j_{\rm 3D}$, while the horizontal red lines show the total $\js$ using the  \Eq{eq:RF12} formula.
This approximation can lead to an over-estimation of the total angular momentum by 10--30\%\ depending on the steepness of the $v(r)$ profile.
}
\label{fig:jprof:example:tng}
\end{figure*}

 \Fig{fig:jprof:example:tng} shows examples of kinematic properties for TNG50 galaxies analyzed in \S~\ref{section:tng50:results}.
 
\section{Redshift evolution of $V/\sigma$}
\label{sec:appendix:VSigma}

\begin{figure*}
\centering
\includegraphics[width=0.85\textwidth]{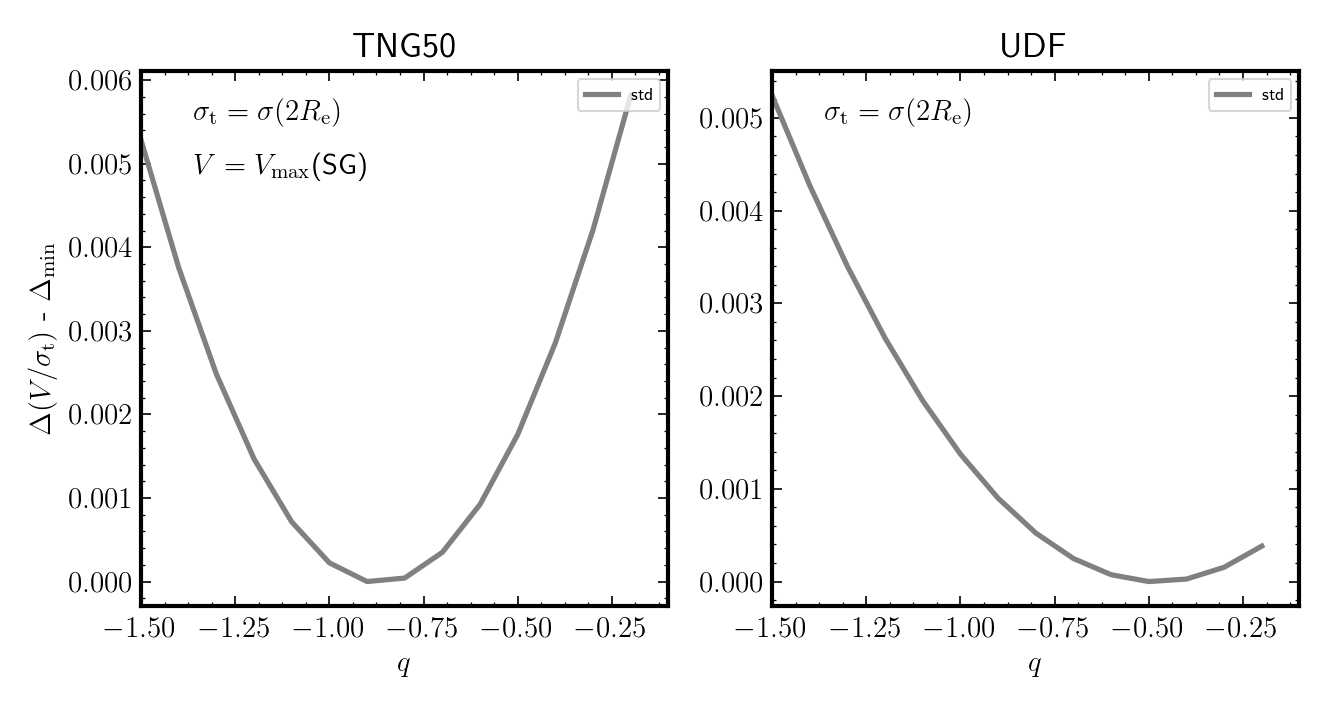}
\caption{{\bf Left:} The $V/\sigma$ standard distribution as a function of the redshift evolution   $q$ (Eq.~\ref{eq:VoverS:evolution}). One sees that $q\sim-0.8$ minimizes the distribution across all redshifts.  {\bf Right:} Same for the UDF sample. The redshift evolution for $V/\sigma$ here is $q\simeq-0.5$.  
\label{fig:appendix:VoverSigma}
}
\end{figure*}

In order to estimate the redshift evolution $q$ of $V/\sigma_{\rm t}$, we minimized the $V/\sigma_{\rm t}$ distribution with respect to $q$.
\Fig{fig:appendix:VoverSigma} shows the results of this excercise for the TNG50 (UDF) sample in the left (right) panel, respectively.
This figure shows that $V/\sigma_{\rm t}$ evolves as $(1+z)^{-0.8}$ in TNG50 and as $(1+z)^{-0.5}$ in the UDF sample.

\end{document}